\documentclass[fleqn,usenatbib]{mnras}

\usepackage{newtxtext,newtxmath}

\usepackage[T1]{fontenc}
\usepackage{ae,aecompl}

\usepackage{graphicx}	
\usepackage{amsmath}	
\usepackage{amssymb}	
\usepackage{hyperref}

\title[High Resolution Molecular Cross Sections]{Molecular Cross Sections for High Resolution Spectroscopy of Super Earths, Warm Neptunes and Hot Jupiters}

\author[S. Gandhi et al.]{
Siddharth Gandhi$^{1,2}$\thanks{E-mail: Siddharth.Gandhi@warwick.ac.uk},
Matteo Brogi$^{1,2,6}$,
Sergei N. Yurchenko$^{3}$,
Jonathan Tennyson$^{3}$,
\newauthor
Phillip A. Coles$^{3}$,
Rebecca K. Webb$^{1}$,
Jayne L. Birkby$^{4}$,
Gloria Guilluy$^{5,6}$,
\newauthor
George A. Hawker$^7$,
Nikku Madhusudhan$^{7}$,
Aldo S. Bonomo$^{6}$,
Alessandro Sozzetti$^{6}$
\\
$^{1}$Department of Physics, University of Warwick, Coventry CV4 7AL, UK\\
$^{2}$Centre for Exoplanets and Habitability, University of Warwick, Gibbet Hill Road, Coventry CV4 7AL, UK\\
$^{3}$Department of Physics and Astronomy, University College London, Gower Street, London WC1E 6BT, UK \\
$^{4}$Anton Pannekoek Institute of Astronomy, University of Amsterdam, Science Park 904, Amsterdam,1098 XH, The Netherlands\\
$^{5}$Dipartimento di Fisica, Universit\`a degli Studi di Torino, via Pietro Giuria 1, I-10125 Torino, Italy\\
$^{6}$INAF - Osservatorio Astrofisico di Torino, Via Osservatorio 20, I-10025 Pino Torinese, Italy\\
$^{7}$Institute of Astronomy, University of Cambridge, Madingley Road, Cambridge, CB3 0HA, UK
}

\date{Accepted XXX. Received YYY; in original form ZZZ}

\pubyear{2020}

\begin{document}
\label{firstpage}
\pagerange{\pageref{firstpage}--\pageref{lastpage}}
\maketitle

\begin{abstract}
High resolution spectroscopy (HRS) has been used to detect a number of species in the atmospheres of hot Jupiters. Key to such detections is accurately and precisely modelled spectra for cross-correlation against the R$\gtrsim$20,000 observations. There is a need for the latest generation of opacities which form the basis for high signal-to-noise detections using such spectra. In this study we present and make publicly available cross sections for six molecular species, H$_2$O, CO, HCN, CH$_4$, NH$_3$ and CO$_2$ using the latest line lists most suitable for low- and high-resolution spectroscopy. We focus on the infrared (0.95-5~$\mu$m) and between 500-1500~K where these species have strong spectral signatures. We generate these cross sections on a grid of pressures and temperatures typical for the photospheres of super Earth, warm Neptunes and hot Jupiters using the latest H$_2$ and He pressure broadening. We highlight the most prominent infrared spectral features by modelling three representative exoplanets, GJ~1214~b, GJ~3470~b and HD~189733~b, which encompass a wide range in temperature, mass and radii. In addition, we verify the line lists for H$_2$O, CO and HCN with previous high resolution observations of hot Jupiters. However, we are unable to detect CH$_4$ with our new cross sections from HRS observations of HD~102195~b. These high accuracy opacities are critical for atmospheric detections with HRS and will be continually updated as new data becomes available.
\end{abstract}

\begin{keywords}
planets and satellites: atmospheres --- methods: data analysis --- techniques: radiative transfer
\end{keywords}



\section{Introduction}

Ground based high resolution Doppler spectroscopy (HRS) has been used to characterise the atmospheres of a growing number of exoplanets in recent years \citep[see e.g. review by][]{birkby2018}. HRS holds strong potential for characterising the atmospheres of exoplanets due to its sensitivity in detecting trace species. A number of chemical species have already been detected on both transiting and non-transiting planets with HRS, most notably nearby hot Jupiters with their strong atmospheric signatures. This has allowed us to expand the chemical inventory of exoplanetary atmospheres. In the future with large ground based telescopes such as ELT, TMT and GMT we will be able to characterise more Earth like planets, with the ultimate goal to observe potential biomarkers \citep{kaltenegger2017, meadows2018}. HRS will thus be key for robust detections of these chemical species and in detecting multiple such biomarkers in the atmosphere.

Numerous species have been discovered in the infrared and the optical with HRS. \citet{snellen2010} first detected molecular absorption due to CO in the infrared in the primary transit of HD~209458~b using the CRIRES spectrograph \citep{kaeufl2004}. Since then we have observed multiple planets with CO absorption in emission spectra \citep[e.g.][]{brogi2012, rodler2012, dekok2013, rodler2013}. H$_2$O has also been regularly detected \citep[e.g.][]{birkby2013, lockwood2014, piskorz2017, birkby2017, sanchez-lopez2019} and there has also recently been evidence for HCN \citep{hawker2018, cabot2019} and CH$_4$ \citep{guilluy2019} in dayside emission spectra. In the optical, high resolution observations have also detected atomic species such as Na \citep{snellen2008, redfield2008, seidel2019} as well as Fe, Ti and other ionic species in the ultra-hot Jupiter KELT-9~b \citep{hoeijmakers2018, hoeijmakers2019}.

In the last few years there has also been significant progress made with observational facilities and atmospheric modelling and inference methods for the observations. In the infrared, instruments such as Keck/NIRSPEC \citep{mclean1998}, CARMENES/CAHA \citep{quirrenbach2014} and GIANO/TNG \citep{oliva2006} have opened up a greater wavelength range at high spectral resolution. Facilities such as iSHELL \citep{rayner2016} and NIRSPEC have spectral ranges extending up to $\sim$5~$\mu$m, beyond which the thermal background becomes significant \citep{birkby2018}. As well as obtaining strong chemical signatures from hot Jupiters \citep[e.g.][]{lockwood2014, piskorz2017, brogi2018, alonso-floriano2019}, such instruments are also capable of probing smaller and cooler planets around bright nearby stars. Planet surveys such as TESS \citep{ricker2015} are capable of providing the most suitable nearby targets for atmospheric follow-up using such facilities.

Developments in atmospheric modelling have also allowed for accurate spectra at high computational efficiency for cross correlation against the high resolution observations \citep[e.g.][]{hawker2018, guilluy2019}. Recently, these developments have been combined with statistical inference tools such as Nested Sampling to allow us to quantify chemical detections of spectroscopically active species. Such high resolution retrievals of the dayside atmosphere of HD~209458~b have retrieved a sub-solar H$_2$O but a solar to super-solar CO abundance \citep{brogi2017, brogi2019, gandhi2019b}. In the near future we will be able to obtain more robust abundance estimates and detection significances with HRS using such frameworks, already achieved regularly with low resolution HST and Spitzer observations \citep[e.g.][]{madhu2009, kreidberg2014, madhu2014, line2016, barstow2017, gandhi2018, mikal-evans2019}.

Central to the detection of chemical species with HRS is accurately and precisely known molecular opacity as a function of frequency \citep{hoeijmakers2015}. \citet{brogi2019} recently demonstrated the influence of the choice of line list on detecting and constraining species such as H$_2$O using HRS. They showed that detections can be missed or result in biased abundance estimates depending on the choice of line list. HRS detections are based on cross correlating numerous molecular spectral lines \citep[see e.g. review by][]{birkby2018} and to maximise this we require accurately known frequencies for each transition line for a given molecule. Well determined frequencies for these lines are therefore vital for reliable chemical detections. Given the influx of data expected over the next few years there is a growing need for a new generation of cross sections for each species for robust and reliable detections using HRS.

In this work we assess the state of the art in molecular cross sections for the most prominent volatile species in super Earths, warm Neptunes and hot Jupiters. We focus on observable chemical signatures in the infrared for H$_2$O, CO, HCN, CH$_4$, NH$_3$ and CO$_2$ from the current generation of high resolution spectrographs. We determine the cross sections in the $\sim$500-1500~K temperature range as in coming years we expect the number of known exoplanets with equilibrium temperatures in this range to increase significantly due to surveys such as TESS \citep{ricker2015}. At such temperatures these molecular species are also expected to be most abundant and therefore have the strongest spectral features from chemical models \citep[see e.g.][]{lodders2002, venot2012, blecic2016, madhu2016, woitke2018}.

We use the latest and most complete line lists for our work. We choose these line lists based primarily on precise line positions making them the most suitable for HRS in the infrared. Our H$_2$O, HCN and NH$_3$ line lists come from the ExoMol database \citep{harris2006, barber2014, polyansky2018,jt771}, the CO and CH$_4$ line list comes from the HITEMP database \citep{rothman2010, li2015, hargreaves2020} and our CO$_2$ line list is obtained from the Ames database \citep{huang2017}. We additionally also compute the H$_2$O cross section using HITEMP to compare the differences between the POKAZATEL line list \citep{polyansky2018}. Each line is spectrally broadened by the temperature and pressure on our grid to determine the cross section, which includes H$_2$ and He pressure broadening from recent work \citep[e.g.][]{faure2013, barton2017, barton2017_broadening}. These line lists are also some of the most complete and are up to date in the spectral range given and therefore also recommended for use in lower resolution equilibrium models and retrievals. Low resolution observations are not able to resolve individual lines, but convolve many lines and detect species through their broad band absorption features. Thus line list completeness ensures that opacity and hence spectral features are not underestimated.

\begin{figure*}
\centering
	\includegraphics[width=\textwidth,trim={0cm 0cm 0cm 0},clip]{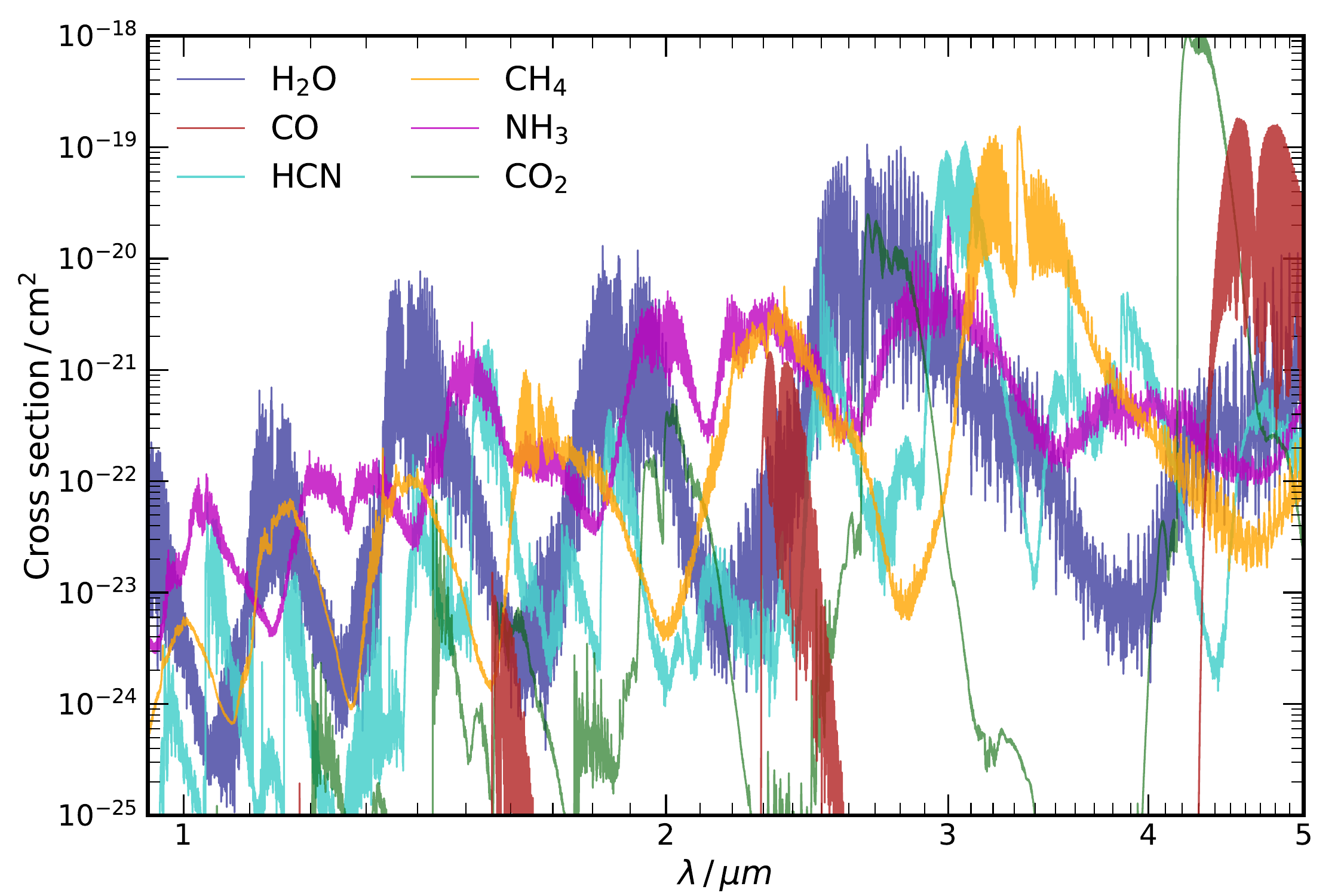}
	\vspace{-5mm}
    \caption{Molecular cross sections for the volatile species discussed in this work. These are shown at a representative temperature of 1000~K and a pressure of 0.1~bar in the 0.95-5~$\mu$m range. The line lists sources for these cross sections are given in Table~\ref{tab:line_list_table}.}     
\label{fig:cs}
\end{figure*}

\begin{table*}
	\centering
	\caption{Line list and pressure broadening references for each of the molecular species.}
	\label{tab:line_list_table}
	\begin{tabular}{l|ll} 
		\hline
		\textbf{Species} & \textbf{Line List} & \textbf{Pressure Broadening due to H$_2$/He} \\
		\hline
		H$_2$O & \cite{polyansky2018} & \cite{petrova2016, barton2017}\\
		CO & \cite{rothman2010,li2015} & \cite{faure2013,gordon2017}\\
		HCN & \cite{harris2006, barber2014} & \cite{mehrotra1985, cohen1973, charron1980}\\
		CH$_4$ & \cite{hargreaves2020} & \cite{varanasi1990, pine1992, gabard2004}\\
		NH$_3$ & \cite{jt771} & \cite{barton2017_broadening}\\
	    CO$_2$ & \cite{huang2013, huang2017} & \cite{padmanabhan2014, gordon2017}\\
		\hline
	\end{tabular}
\end{table*}

In what follows, we discuss each line list and the computation of the cross sections in section \ref{sec:cross_secs}. In section \ref{sec:spectra} we use the cross sections for each species to generate high resolution spectra for three known exoplanets in transmission or emission and explore the molecular features of each species in their spectra. We then go on in section \ref{sec:prev_detections} to explore previous HRS detections of H$_2$O, CO, HCN and CH$_4$ in five hot Jupiters and compare and contrast the new cross sections with previous work. Finally we present the discussion and conclusions in section \ref{sec:conclusion}.

While the line lists discussed here are currently the most suitable for HRS, they will be continually updated once more complete and accurate line list data becomes available. We make the most up-to-date data publicly available on the \href{ https://osf.io/mgnw5/?view_only=5d58b814328e4600862ccfae4720acc3}{Open Science Framework (OSF)}\footnote{\url{ https://osf.io/mgnw5/?view_only=5d58b814328e4600862ccfae4720acc3}}.

\section{Molecular Cross Sections}\label{sec:cross_secs}

\subsection{Line Lists}

The line lists for each molecular species have been chosen to be the most accurate and up to date for temperatures in the range $\sim$500-1500~K typical of the hot Jupiters, warm Neptunes and super Earths we are likely to observe and where they are most prominent. Where available these have been empirically determined to ensure more accurate transition frequencies for high resolution applications.

\begin{figure*}
\centering
	\includegraphics[width=\textwidth,trim={0cm 0.0cm 0cm 0},clip]{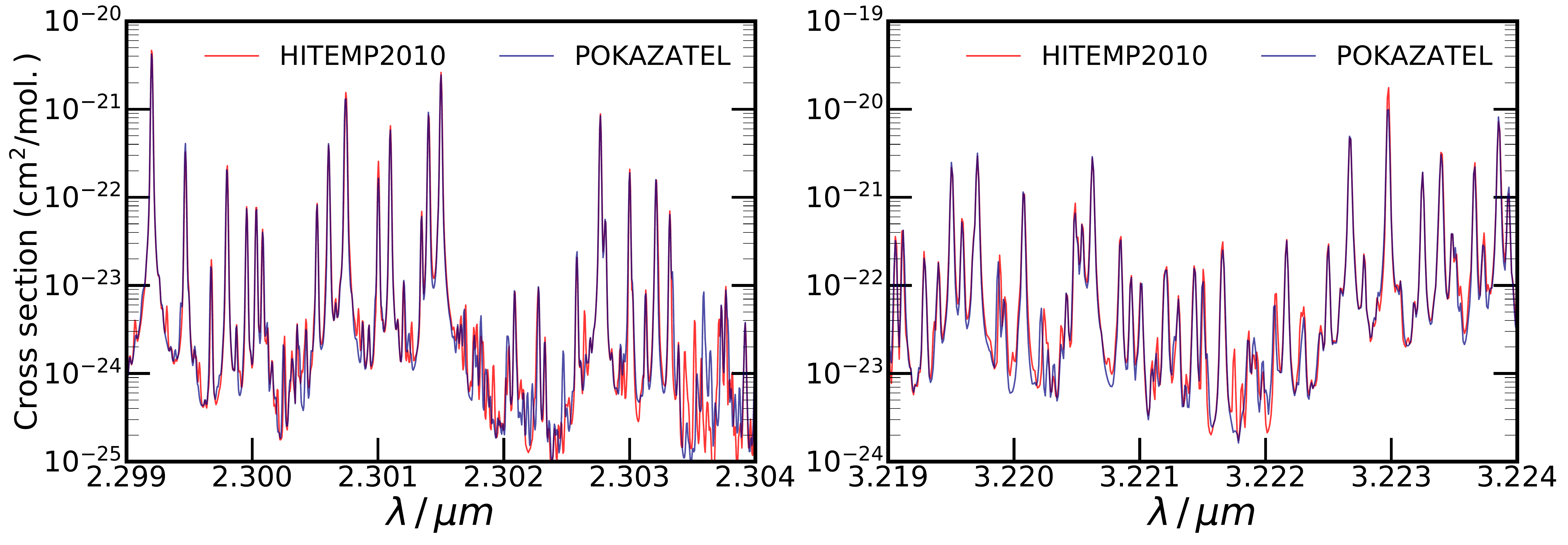}
	\vspace{-5mm}
    \caption{Comparison of the cross sections generated from the H$_2$O POKAZATEL \citep{polyansky2018} and HITEMP \citep{rothman2010} line lists. These are generated at 0.01~cm$^{-1}$ spacing and shown in the 2.3~$\mu$m (at R$\sim$430,000) and 3.2~$\mu$m band (at R$\sim$310,000), where H$_2$O has been observed with the CRIRES spectrograph \citep[e.g.][]{brogi2013, brogi2014, birkby2013, birkby2017, hawker2018, cabot2019}. These cross sections are given at a representative temperature of 1000~K and a pressure of 0.1~bar.}
\label{fig:cs_h2o_compare}
\end{figure*}

\subsubsection{H$_2$O}

H$_2$O is one of the most well studied molecules because of its importance and detectability in exoplanetary atmospheres \citep[e.g. see review by][]{madhu2019}. It has been detected with numerous high resolution observations and is one of the dominant sources of opacity for all planets over the temperature range considered here. H$_2$O is of particular interest for HRS given that observations in some spectral ranges show high signal-to-noise detections whereas other wavelength ranges show absent or weak features for the same exoplanets. For instance, H$_2$O was detected in the $\sim$3-3.5~$\mu$m range for for $\tau$~Bo{\"o}tis~b \citep{lockwood2014} but was undetected in the $\sim$2.3~$\mu$m range observations in \citet{brogi2012}, despite H$_2$O showing strong opacity in both ranges (see Figure~\ref{fig:cs}). To compute our cross sections we use the POKAZATEL line list from ExoMol \citep{polyansky2018} which uses ab initio calculations for a fully complete line list up to dissociation. In addition, empirical energy levels \citep{jt539} determined with the MARVEL \citep[Measured Active Rotation-Vibration Energy Levels,][]{jt412} procedure allow for accurate line positions, particularly for the strongest lines which are the most well measured and easiest to detect with HRS. The cross section for H$_2$O at representative photosphere conditions is shown in Figure \ref{fig:cs}. It clearly shows strong cross section bands in the $\sim$1.4~$\mu$m, $\sim$1.9~$\mu$m and $\sim$2.7~$\mu$m spectral range.

We compared the POKAZATEL line list to the HITEMP line list \citep{rothman2010} at a representative temperature and pressure for the 2.3~$\mu$m and the 3.2~$\mu$m range as shown in Figure~\ref{fig:cs_h2o_compare}. The strongest lines show excellent agreement for both line strengths and the transition frequencies. The weaker lines do show some differences, but these are minor and the spectra generated with each remain very similar as they are dominated by the strongest of the lines. Cross correlation with observed spectra also therefore do not show any significant differences between the line lists as discussed further in section \ref{sec:h2o_hd189}. 

As the temperature exceeds $\sim$1200~K, the difference between the two line lists does increase slightly. This is primarily in the spectral regions where the cross section is weaker, such as $\sim$2.2~$\mu$m and $\sim$4~$\mu$m (see Figure \ref{fig:cs}). This is not unexpected given that experimental verification of the lines at such wavelengths is difficult due to the lower opacity. However, there is a need for accurate line positions for these weaker lines given that high resolution ground based spectroscopy is most accurate in less contaminated regions where telluric absorption from H$_2$O in the Earth's atmosphere is weaker.

\subsubsection{CO}

CO is another species which along with H$_2$O has been detected in numerous transiting and non-transiting hot Jupiters with HRS. We use the HITEMP CO line list for our work \citep{rothman2010,li2015}. The cross section of CO is shown in Figure \ref{fig:cs}. This clearly shows distinct bands with a strong cross section at $\sim$1.6~$\mu$m, $\sim$2.3~$\mu$m and $\sim$4.6~$\mu$m. Observations in the 2.3~$\mu$m range were used to detect CO in the transmission spectrum of HD~209458~b \citep{snellen2010} and also provide numerous high signal-to-noise detections in emission spectra \citep[e.g.][]{brogi2012, rodler2012, dekok2013, rodler2013}. The updated HITEMP line list offers more complete coverage at lower wavelengths ($\lesssim$1.6~$\mu$m). This allows us to explore CO at the shorter wavelengths where the ground based HRS observations are able to probe, although it is likely that CO features will be weaker than other species. H$_2$ and He broadening coefficients for CO are provided in \citet{faure2013} and \citet{gordon2017} as shown in table \ref{tab:line_list_table}. The pressure broadening is unlikely to have a significant impact on HRS detections given that these observations are more sensitive to the line centres at low pressures ($\lesssim$10$^{-2}$~bar). However, we include this for completeness and to ensure accuracy for lower resolution observations which are more sensitive to the line wings. At such pressures the broadening can become significant \citep{hedges2016}.

\subsubsection{HCN}

We use the ExoMol line list for HCN \citep{harris2006, barber2014}, a high accuracy line list developed for HCN and HNC which has been verified against experimental measurements \citep{mellau2011}. We compared the ExoMol line list to HITRAN \citep{rothman2013, gordon2017} at room temperature and also found excellent agreement. The more complete coverage of this line list means that the cross sections begin to deviate at higher temperatures where other lines increase in strength and thus become more significant. HCN shows significant cross section at $\sim$3.2~$\mu$m (see Figure \ref{fig:cs}) and two hot Jupiters, HD~209458~b and HD~189733~b, have shown evidence for the species \citep{hawker2018, cabot2019} in their emission spectra using the ExoMol line list in this spectral range. HCN is expected to be a significant source of opacity for hot Jupiter atmospheres with temperatures $\gtrsim$1500~K when the atmospheric C/O ratio is super-solar \citep{madhu2012, moses2013, molliere2015, drummond2019}.

\subsubsection{CH$_4$}

Our CH$_4$ cross section is calculated from the new HITEMP line list \citep{hargreaves2020}. This new addition to the HITEMP database utilises ab initio calculations with empirical corrections \citep{rey2017} as well as the HITRAN2016 database \citep{gordon2017} to produce a high temperature line list up to 2000~K. \citet{hargreaves2020} found good agreement with experimentally measured opacity at high temperature from \citet{hargreaves2015} and \citet{wong2019}. The HITEMP line list uses effective lines allowing for efficient computation of opacities for such a large molecule with many billions of transitions. We obtain the H$_2$ and He pressure broadening coefficients from ExoMol, adopting the a0 coefficients for the present work from the references shown in Table~\ref{tab:line_list_table}. These H$_2$ and He broadening coefficients are discussed further in section \ref{sec:broadening_coeffs}. Recently, high temperature H$_2$ broadening coefficients have been measured for CH$_4$ in the 2840-3000~cm$^{-1}$ range \citep{gharib-nezhad2019}. Such coefficients will be vital to accurately constrain CH$_4$ as we observe cooler exoplanet atmospheres as chemical models have shown that it is expected to be the dominant carbon bearing species below $\sim$1000~K \citep{madhu2012, moses2013, drummond2019}. We will therefore constantly update the cross sections to provide the most accurate for both low- and high-resolution spectroscopy of exoplanet atmospheres.

\subsubsection{NH$_3$}

The NH$_3$ line list comes from ExoMol \citep{jt771} and covers the whole infrared range considered in this work (0.95-5~$\mu$m). This line list has also been corrected using empirical energy levels \citep{jt608,jtNH3update} up to 6000~cm$^{-1}$ ($\gtrsim$1.66~$\mu$m) generated using the MARVEL procedure and is hence ideal for HRS. Numerous strong features can be seen in the cross section in Figure~\ref{fig:cs} at $\sim$1.6~$\mu$m (H band) and $\sim$2.2~$\mu$m (K band). It may potentially be detectable with low resolution observations of hot Jupiters if present at sufficient abundance \citep{macdonald2017}. The higher sensitivity of HRS to trace species mean that the high accuracy ExoMol line list may be key to robust constraints of NH$_3$ in the future. In addition, the latest generation of high resolution spectrographs may also be able to explore cooler planets with temperatures $\lesssim$1000~K where NH$_3$ may be more abundant with $\log(X_\mathrm{NH_3})\gtrsim-5$ \citep{moses2013_gj436}. 

\subsubsection{CO$_2$}

The Ames line list for CO$_2$ \citep{huang2013, huang2017} offers accurate line positions and transition strengths with a root mean square deviation of $\lesssim$0.02~cm$^{-1}$ in line position from experimentally determined values. This accuracy is required for HRS given the high resolution observations. Tentative constraints of CO$_2$ have been observed in previous low resolution and high resolution observations \citep{stevenson2010, birkby2013}, but further observations are needed to confirm the presence of the species. CO$_2$ is expected to become prominent at cooler temperatures $\lesssim$1000~K. This is particularly so as we explore non H$_2$-rich atmospheres, where it is expected to be highly abundant in the atmosphere for metallicities $\gtrsim$1000$\times$ solar \citep[e.g.][]{moses2013_gj436}. 

\subsection{Determining the Cross Section}

\begin{table}
\begin{center}
\caption{The temperature and pressure grid for our cross sections. The cross sections are generated at each pressure and temperature at 0.01~cm$^{-1}$ wavenumber spacing in the 0.95-5~$\mu$m range (10,526-2000~cm$^{-1}$).}\label{table:cross_sec_grid}
\begin{tabular}{ |c| c c c c c c c|} 
\hline
 \textbf{T(K)} & 400& 600&800&1000&1200&1400&1600 \\ 
 \hline
 \textbf{P(bar)} & $10^{-5}$ & $10^{-4}$ & $10^{-3}$ & $10^{-2}$ & $10^{-1}$&1&10\\ 
 \hline
\end{tabular}
\end{center}
\end{table}

We now discuss how we calculate the cross section from the line list of each molecular species. Table~\ref{table:cross_sec_grid} shows the grid that we compute the cross sections on, ranging between 400-1600~K in temperature and 10$^{-5}$-10~bar in pressure. At each pressure and temperature the lines are spectrally broadened into a Voigt profile according to the method in \citet{gandhi2017}. 
We compute the cross sections in the infrared on a uniform grid of 0.01~cm$^{-1}$ wavenumber spacing between 10,526-2000~cm$^{-1}$ (0.95-5~$\mu$m). This corresponds to a spectral resolution of $\mathrm{R} = 10^6$ at 1~$\mu$m.
To compute the pressure broadening we use the latest H$_2$ and He broadening coefficients obtained from sources listed in Table~\ref{tab:line_list_table}. In addition, we also include the natural broadening of each line which arises from the uncertainty principle \citep{gray1976}. The summed contribution from each line in the line list gives the overall cross section for each species. We first describe how to compute the line strengths at each temperature, followed by the broadening procedure.

\subsubsection{Line Strengths}

As the temperature is changed the populations within each of the states are altered and hence the line strengths of each line in a line list, $S(T)$, also vary. These line strengths are often provided in the line lists relative to a reference temperature (e.g. 296~K). The line strength at a general temperature $T$ is then calculated from these by \citep{gordon2017},
\begin{align}
S(T) &= S_0\frac{Q(T_\mathrm{ref})}{Q(T)}\frac{\mathrm{exp}(-E_\mathrm{lower}/k_bT)}{\mathrm{exp}(-E_\mathrm{lower}/k_bT_{\mathrm{ref}})}\frac{1-\mathrm{exp}(-\nu_0/k_bT)}{1-\mathrm{exp}(-\nu_0/k_bT_{\mathrm{ref}})},
\end{align}
where $S_0$ is the line strength at the reference temperature $T_\mathrm{ref}$. $E_\mathrm{lower}$ is the lower energy state of the transition (in cm$^{-1}$) and $\nu_0$ is the frequency of the transition ($E_\mathrm{upper}-E_\mathrm{lower} = \nu_0$). The partition function $Q$ is
\begin{align}
Q(T) &= \sum_j g_j \mathrm{exp}(-E_j/k_bT)
\end{align}
with the degeneracy of the state $j$ given by $g_j$ and where $k_b$ is the Boltzmann constant (in cm$^{-1}$ K$^{-1}$). The line strength $S(T)$ can also be calculated directly through the Einstein coefficient $A$ of a transition, 
\begin{align}
S(T) &= \frac{A g}{8\pi c \nu_0^2}\frac{\mathrm{exp}(-E_\mathrm{lower}/k_bT)}{Q(T)} (1-\mathrm{exp}(-\nu_0/k_bT)).
\end{align}
We must calculate the line strength of each line at each temperature that the cross sections are computed.

\subsubsection{Thermal and Pressure Broadening}

Every line is broadened according to the temperature and pressure in order to determine the cross section as a function of frequency. The broadening due to the temperature is caused by the velocity distribution of the molecules in the gas which Doppler shifts the transition frequency according to the distribution of speeds. This results in each transition line being spread over frequency. In this case the thermal broadening results in a Gaussian line profile \citep{hill2013}, given by
\begin{align}
f_G(\nu-\nu_0) &= \frac{1}{\gamma_G \sqrt[]{\pi}}\mathrm{exp}\left(-\frac{(\nu-\nu_0)^2}{\gamma_G^2}\right),\\
\gamma_G &= \sqrt[]{\frac{2k_bT}{m}}\frac{\nu_0}{c},
\end{align}
where $\nu$ and $\nu_0$ represent the frequency and the line transition frequency respectively (in cm$^{-1}$). Here, $m$ represents the mass (in kg) of the molecule and $T$ represents the temperature (in K). 

Collisions with other molecules at pressure alters each molecule's state decay times and results in a Lorentzian profile with frequency. In addition to this, each line is also naturally broadened due the uncertainty principle \citep{gray1976}. This is often much weaker than pressure broadening, particularly in the photospheres of exoplanets in our HRS observations, but we include this effect for completeness. The combined Lorentzian profile from pressure and natural broadening is given by
\begin{align}
f_L(\nu-\nu_0) &= \frac{1}{\pi}\frac{\gamma_L}{(\nu-\nu_0)^2+\gamma_L^2},\\
\gamma_L &= P \sum_i \left(\frac{T_{\mathrm{ref}}}{T}\right)^{n_i} \gamma_{i}\, X_i + \gamma_N,\\
\gamma_N &= 0.22 \times 10^{-2}\frac{\nu_0^2}{4\pi c}
\end{align}
where $\gamma_L$ and $\gamma_N$ are Lorentzian and natural values of the corresponding half-widths-at-half-maximum (HWHM) and $X_i$ is the mixing ratio of a specific broadening species $i$. The pressure broadening coefficients $\gamma_{i}$ and $n_i$ represent the Lorentzian HWHM and the power law for the temperature respectively. A reference temperature $T_\mathrm{ref}$ of 296~K is most commonly used. In our work the pressure broadening is included from from $i=\mathrm{H_2}$ and $i=\mathrm{He}$ (Table~\ref{tab:line_list_table}), with the mixing ratio $X_i$ set from solar abundances \citep{asplund2009}.

Natural broadening arises as a result of the finite lifetime of a state which results in a $\Delta \nu$ from the uncertainty principle. The 0.22$\times10^{-2}$ factor in the natural broadening width has been derived for hydrogen \citep{gray1976} but we use this value for all of the species in our work given the absence of any other calculated values. The value of $\gamma_N$ for Sodium with experimental measurements of one of the the Na D line transitions at $\sim$0.589~$\mu$m \citep{bernath2015} does show good agreement. The natural broadening is relatively weak compared to thermal and pressure broadening unless considering low pressures and temperatures and high wavenumbers so any differences that may arise for species in our work are also less of a concern.

The full broadening profile is a convolution of the Gaussian from thermal broadening and the Lorentzian profile. This is known as a Voigt function,
\begin{align}
f_V(\nu-\nu_0) &= \int_{-\infty}^{\infty}f_G(\nu'-\nu_0)f_L(\nu-\nu')d\nu'.
\end{align}
Defining 
\begin{align}
u&\equiv\frac{\nu-\nu_0}{\gamma_G},\\
a&\equiv\frac{\gamma_L}{\gamma_G},
\end{align}
the Voigt function can be cast in terms of the real part of the normalised Faddeeva function, $w(x + iy)$, to
\begin{align}
f_V(\nu - \nu_0,\gamma_L,\gamma_G) &= \frac{\mathrm{Re}(w(u+ia))}{\gamma_G \sqrt[]{\pi}}.
\end{align}
The cross section $\sigma_\nu$ at a frequency $\nu$ for a transition line with strength $S(T)$ is then given by \citep{gandhi2017}
\begin{align}
\sigma_\nu = S(T)f_V(\nu- \nu_0,\gamma_L,\gamma_G).
\end{align}

The temperature and pressure grid used to compute the cross sections is given in Table~\ref{table:cross_sec_grid} and encompasses typical ranges likely in the photosphere. We have ignored the frequency shifts of the line positions with pressure for all species except CO as these are negligible for our work. 

\subsubsection{Pressure Broadening Coefficients}\label{sec:broadening_coeffs}

Table~\ref{tab:line_list_table} shows the references for the broadening coefficients $n_i$ and $\gamma_i$ for each of the volatile species. For each species we use H$_2$ and He broadening to work out the cross section as a function of wavelength for each pressure and temperature. We calculate the broadened lines out to 500 Voigt widths according to previous works \citet{hedges2016, gandhi2017}. This corresponds to a minimum line extent of $\sim$20~cm$^{-1}$. The broadening coefficient files are provided on the ExoMol database\footnote{\url{http://exomol.com/data/data-types/broadening_coefficients/}}. The coefficients are determined by the J-quantum numbers of each transition. We adopt the a0 coefficients for our present work which requires only the lower state J-quantum number. We then determine the overall $n_i$ and $\gamma_i$ by summing over the H$_2$ and He broadening coefficients weighted by their abundance. However for CO$_2$ these coefficients were not available so we adopt $\gamma_i = 0.11$~cm$^{-1}$/atm from \citet{padmanabhan2014} and $n_i$ from HITRAN \citep{gordon2017}.

\begin{figure*}
\centering
	\includegraphics[width=\textwidth,trim={0cm 0cm 0cm 0},clip]{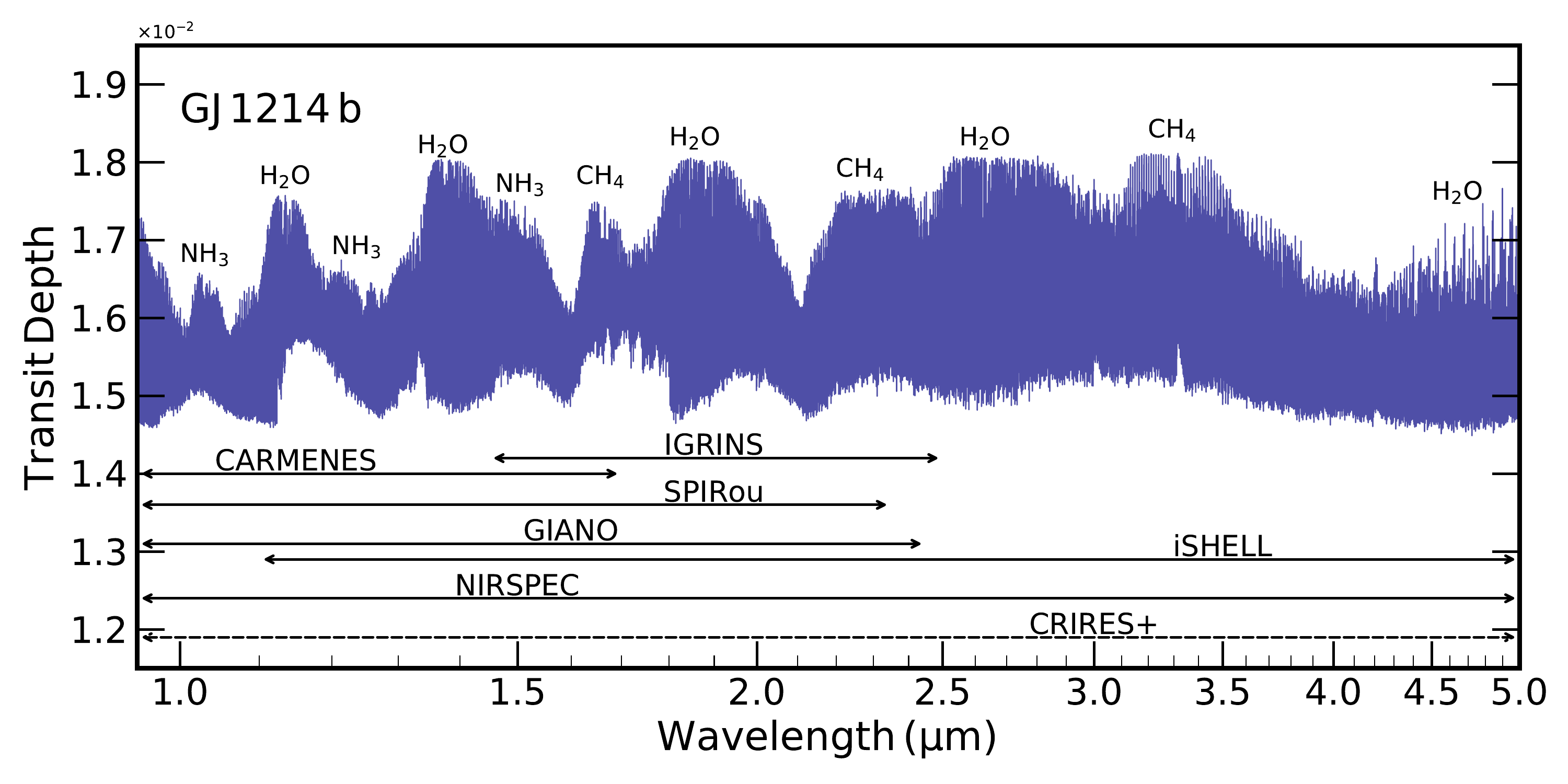}
	\vspace{-5mm}
    \caption{The cloud free transmission spectrum of GJ~1214~b showing the transit depth with wavelength. We also show the spectral ranges for the current generation of high resolution spectrographs. The upcoming CRIRES+ on the VLT has been marked with a dashed line as the exact formatting of the orders has not yet been determined, hence we show the full observable range.} 
\label{fig:gj1214_spec}
\end{figure*}

\section{Model Spectra}\label{sec:spectra}

We will now use the cross sections generated from the line lists as discussed in section \ref{sec:cross_secs} to accurately model the spectra of known super Earths, warm Neptunes and hot Jupiters. We model three exoplanets with a range in mass, radius and equilibrium temperature to explore the spectral features of each species in the infrared spectrum. Most current observations target hot Jupiters, and therefore do not encompass the full range of systems observable today. Surveys such as TESS \citep{ricker2015} will find more warm Neptunes and super Earths suitable for characterisation with HRS and hence we also model the spectra of these cooler targets. We generate spectra using GENESIS \citep{gandhi2017} for emission and AURA \citep{pinhas2018} for transmission. The spectra have been generated to match the cross section grid at a wavenumber spacing of 0.01~cm$^{-1}$ in the 0.95-5~$\mu$m range. In addition to the opacity from the molecular species, we also include collisionally induced absorption (CIA) from H$_2$-H$_2$ and H$_2$-He interactions \citep{richard2012}. 

We begin with the super Earth/sub-Neptune GJ~1214~b, and then discuss the warm sub-Neptune GJ~3470~b. In spite of the observational evidence for clouds in these two exoplanets, in this work we model the cloud free spectrum as a representative case for each exoplanet class. Both of these are modelled under transmission geometries given that the lower temperatures and smaller radius make thermal emission more difficult to detect. Hence transmission spectroscopy of super Earths and warm Neptunes will likely produce the strongest detections and constraints on the volatile molecular species. In addition, we also model the dayside emission spectrum of HD~189733~b, a hot Jupiter, which has had H$_2$O, CO and HCN detected in the dayside atmosphere \citep{birkby2013, rodler2013, cabot2019}.

\begin{figure*}
\centering
	\includegraphics[width=\textwidth,trim={0cm 0cm 0cm 0},clip]{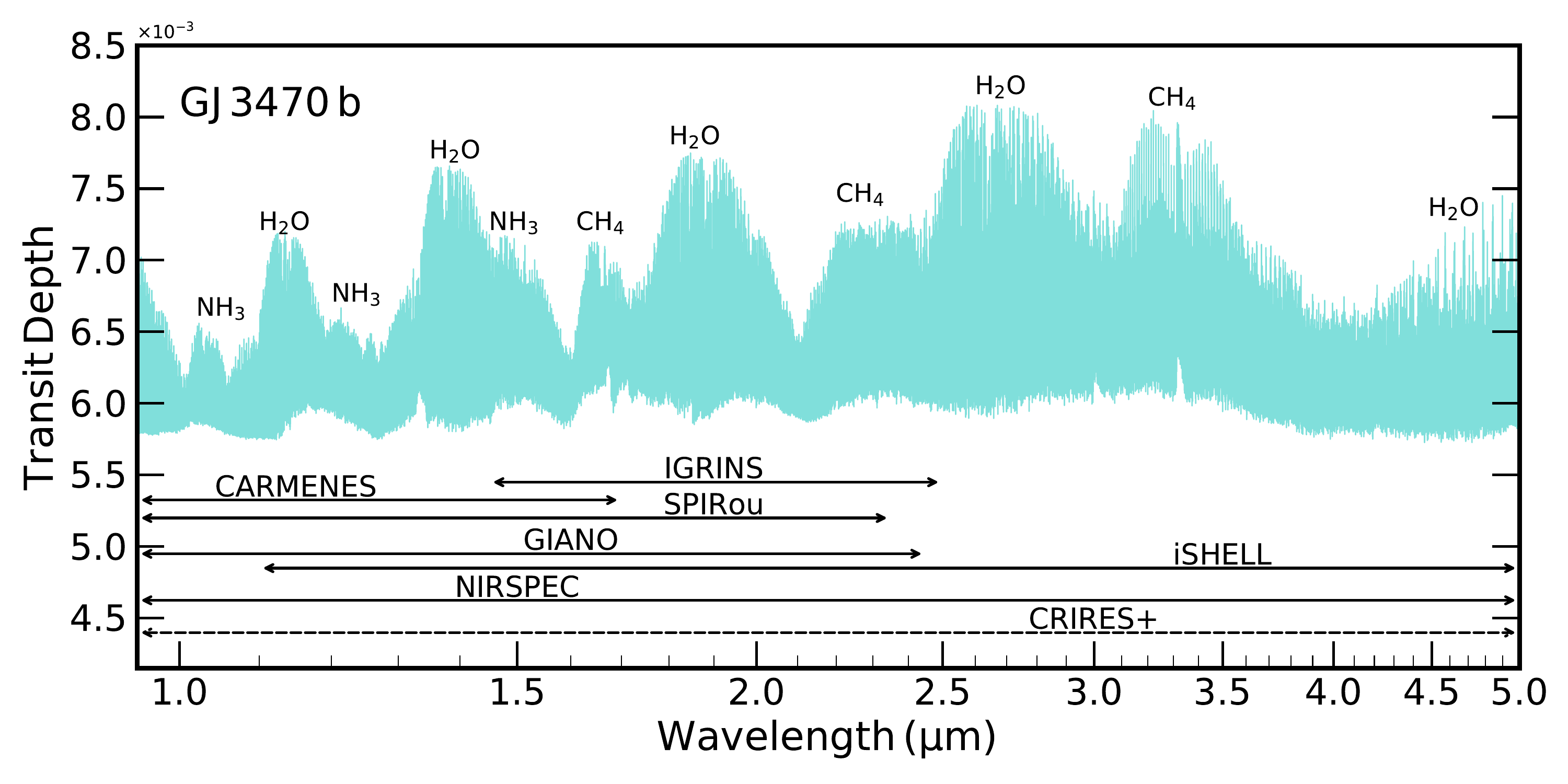}
	\vspace{-5mm}
    \caption{The cloud free transmission spectrum of GJ~3470~b showing the transit depth with wavelength. We also show the spectral ranges for the current generation of high resolution spectrographs. The upcoming CRIRES+ on the VLT has been marked with a dashed line as the exact formatting of the orders has not yet been determined, hence we show the full observable range.}     
\label{fig:gj3470_spec}
\end{figure*}

\begin{figure*}
\centering
	\includegraphics[width=\textwidth,trim={0cm 0cm 0cm 0},clip]{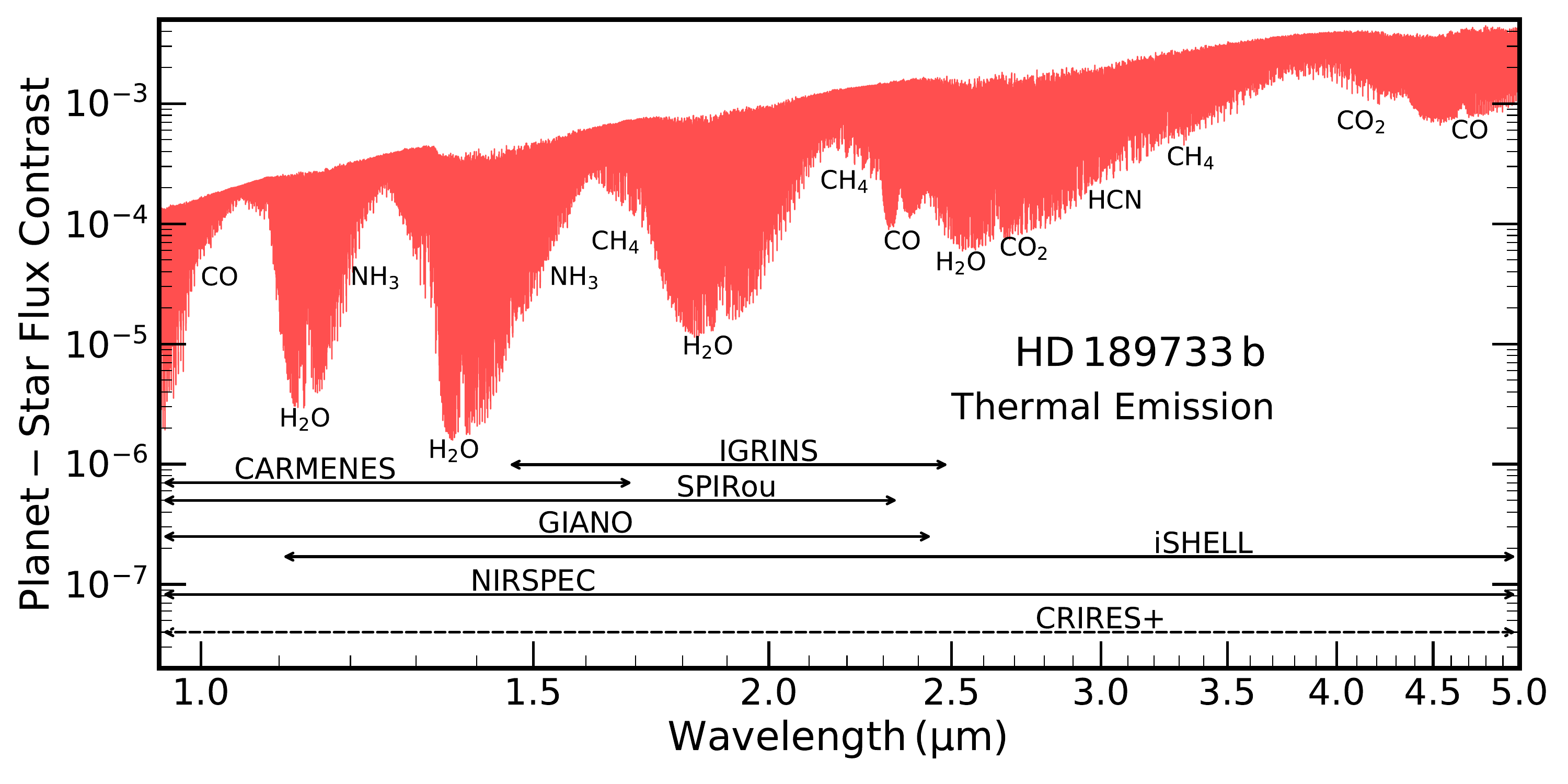}
	\vspace{-5mm}
    \caption{Planet-star flux ratio in the thermal emission spectrum of HD~189733~b. We model a radiative-convective equilibrium atmosphere \citep{gandhi2017} with the volatile molecular species in chemical equilibrium at solar abundance. We also show the spectral ranges for the current generation of high resolution spectrographs. The upcoming CRIRES+ on the VLT has been marked with a dashed line as the exact formatting of the orders has not yet been determined, hence we show the full observable range.}     
\label{fig:hd189_spec}
\end{figure*}

\subsection{Super Earth/Sub-Neptune: GJ~1214~b}

The exoplanet GJ~1214~b has a radius of $\sim$2.6~R$_\oplus$ and a mass of $\sim$6.5~M$_\oplus$ and is one of the coolest exoplanets observed thus far with an equilibrium temperature of $\sim$550~K \citep{carter2011}. We model the cloud free spectrum at 100$\times$ solar abundance, with $\log(\mathrm{H_2O}) = -1.2$, $\log(\mathrm{CH_4}) = -1.5$ and $\log(\mathrm{NH_3}) = -2$. This is shown in Figure \ref{fig:gj1214_spec}. This shows a highly varied transit depth with wavelength due to opacity from the molecular species present in the atmosphere. The broad molecular absorption bands from the cross section of each of the species result in the planet's atmosphere becoming opaque at higher altitudes. This results in a larger effective planet radius and hence transit depth. Molecular features which result form the cross section of each species can clearly be seen in the spectrum. The molecule corresponding to each spectral feature is highlighted in the spectrum in Figure \ref{fig:gj1214_spec}.

Numerous strong spectral lines can be seen for H$_2$O, CH$_4$ and NH$_3$ for cross correlation in the infrared. The spectrum is dominated by the absorption of H$_2$O due to its large cross section and abundance. CH$_4$ also shows some strong features, particularly in the $\sim$1.8~$\mu$m and $\sim$3.5~$\mu$m range where it has a strong cross section. The NH$_3$ on the other hand has generally weaker spectral signatures than the H$_2$O and CH$_4$ due to the lower abundance. However, features can clearly be seen in the spectral ranges where H$_2$O and CH$_4$ opacity is weak. We have modelled the cloud free atmosphere to demonstrate the new high resolution cross sections in this work but we note that GJ~1214~b has shown evidence for a high altitude cloud deck \citep{bean2010, berta2012, kreidberg2014_gj1214}. Cloudy atmospheres reduce the extent of spectral features, making detections more difficult due to the shallower spectral lines for each species. Understanding how cloudy exoplanets affect high resolution spectroscopy is thus extremely important for robust detections and will be extensively treated in the future but is beyond the scope of our current work.

\subsection{Warm Neptune: GJ~3470~b}

The exoplanet GJ~3470~b is more massive than GJ~1214~b with a mass of $\sim$12~M$_\oplus$ and thus falls into the category of warm sub-Neptunes. This planet also has a higher equilibrium temperature of $\sim$650~K. The atmosphere of GJ~3470~b has recently been observed with HST WFC3, HST STIS and Spitzer spectrographs and shown evidence for H$_2$O in the atmosphere at solar abundance \citep{benneke2019}. The cloud free transmission spectrum is shown in Figure \ref{fig:gj3470_spec}. Despite the larger planet radius, the transit depth is lower than for GJ~1214~b due to the larger stellar radius. Here, we have assumed a temperature profile based on the best fit retrieval in \citet{benneke2019}, which indicated that the upper layers of the atmosphere were at $\sim$600~K, consistent with the equilibrium temperature. We have modelled this atmosphere assuming a solar abundance of the volatile species, with $\log(\mathrm{H_2O}) = -3.2$, $\log(\mathrm{CH_4}) = -3.5$ and $\log(\mathrm{NH_3}) = -4$. Numerous spectral features can be seen for the volatile species throughout the infrared. As with GJ~1214~b, the atmosphere is dominated by the strong absorption from H$_2$O and CH$_4$, with weaker features from NH$_3$. The spectral features are generally weaker than for GJ~1214~b due to the lower atmospheric abundance of each species.

\subsection{Hot Jupiter: HD~189733~b}

The hot Jupiter HD~189733~b is one of the most well observed exoplanets in both low resolution \citep{jt400,crouzet2014} and high resolution under emission given its strong signature from its extended hot H$_2$-rich atmosphere. Previous high resolution observations of the dayside with CRIRES, NIRSPEC and CARMENES have shown evidence for H$_2$O \citep{birkby2013, alonso-floriano2019}, CO \citep{rodler2013, dekok2013} and HCN \citep{cabot2019}. We model the dayside atmosphere of this hot Jupiter in chemical and radiative-convective equilibrium assuming a non-inverted pressure-temperature profile \citep{gandhi2017}, consistent with the observational constraints.

Figure~\ref{fig:hd189_spec} shows the planet-star flux ratio for HD~189733~b. This shows that at such high temperatures ($\sim$1400~K) the spectrum is dominated by the presence of H$_2$O and CO. CH$_4$ and NH$_3$ are now at lower abundances in the atmosphere due to the higher temperature. Hence only small absorption features can be seen from CH$_4$, NH$_3$, CO$_2$ and HCN and only in the spectral ranges where the H$_2$O and CO absorption is weak. However, the current generation of spectrographs are still be able to detect some of these trace species \citep{cabot2019} in the atmosphere due to the strong emission from the planet, particularly if the atmospheric C/O ratio is super-solar to potentially enhance their abundance \citep{madhu2012, moses2013, drummond2019}.

\begin{figure*}
\centering
    \includegraphics[width=0.42\textwidth,trim={0.2cm 0cm 3.6cm 0},clip]{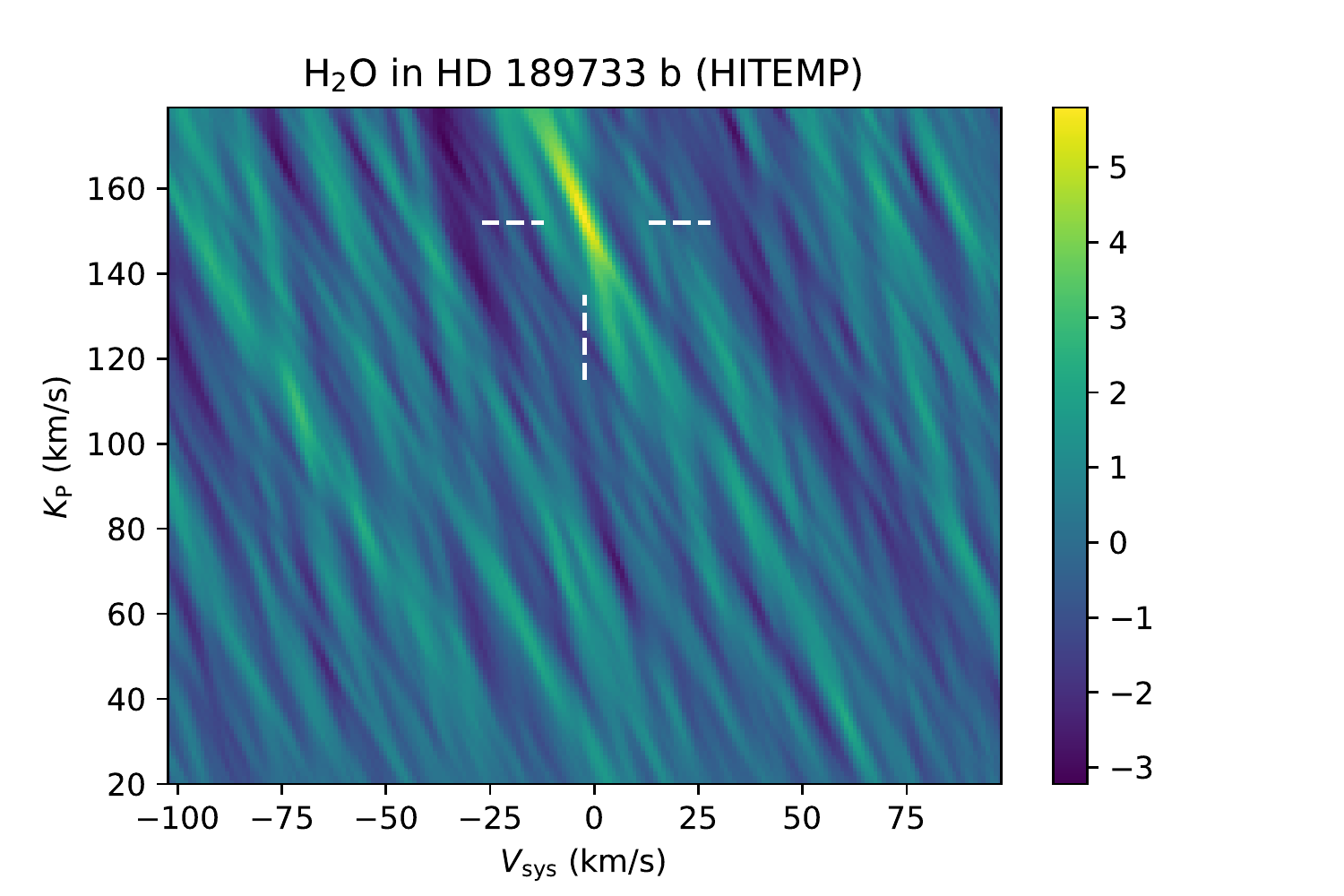}
	\includegraphics[width=0.42\textwidth,trim={1.8cm 0cm 2.0cm 0},clip]{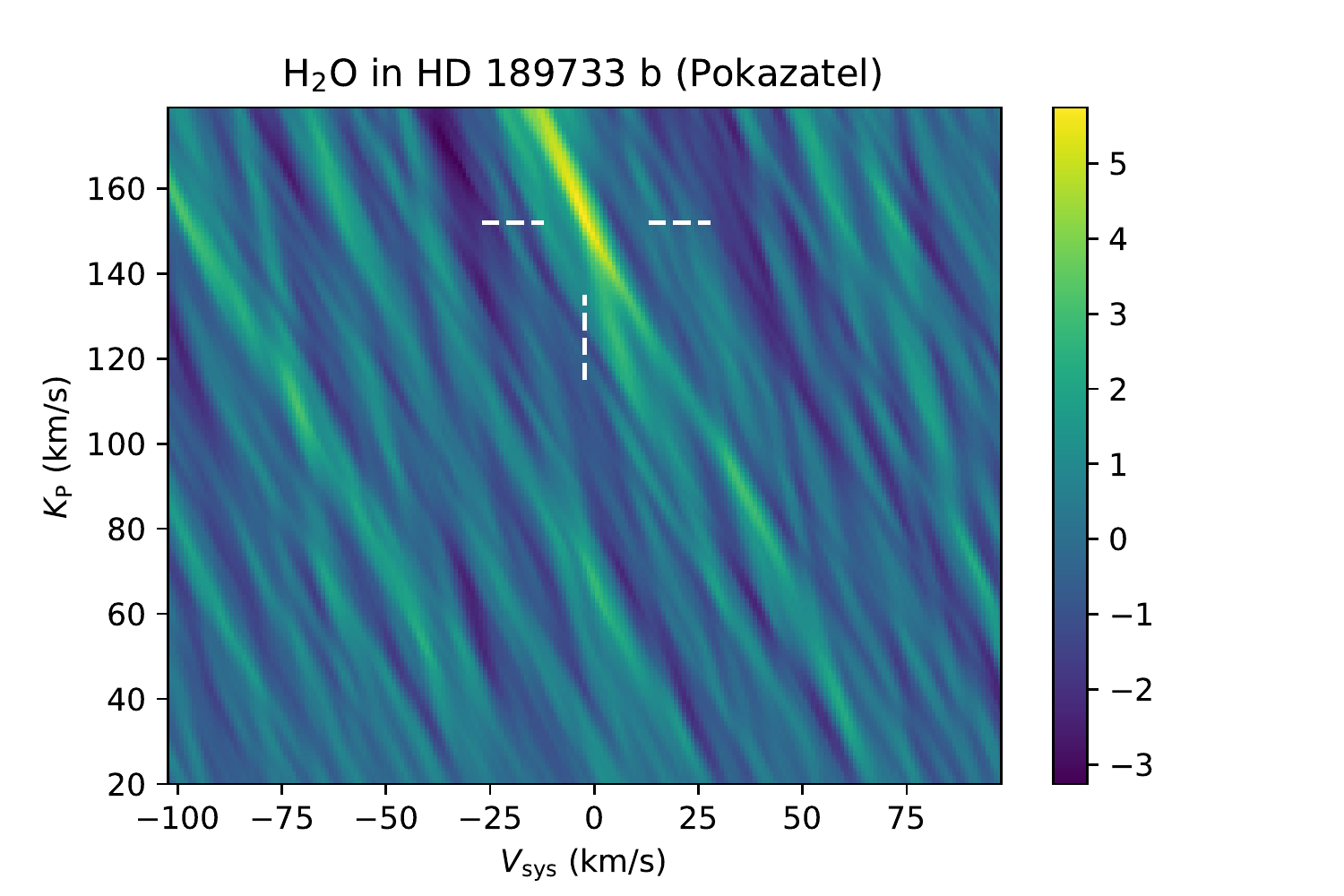}
    \caption{H$_2$O signal obtained by cross correlating VLT/CRIRES spectra of exoplanet HD~189733~b around 3.2 $\mu$m with a best-fitting model matching the parameters of \citet{birkby2013} and using either opacities from \citet{rothman2010} (left panel) or the newest opacities from \citet{polyansky2018} (right panel). The cross correlation signal is shown as function of systemic velocity $V_\mathrm{sys}$ and maximum orbital radial velocity $K_\mathrm{P}$ and shows excellent agreement between the two models, both in terms of localisation and strength. The known values of (-2.4, 151) km s$^{-1}$ are marked with dashed white lines and are a good match to these observations.}
\label{fig:hd189733_ccf}
\end{figure*}

\begin{figure*}
\centering
    \includegraphics[width=0.42\textwidth,trim={0.2cm 0cm 3.6cm 0},clip]{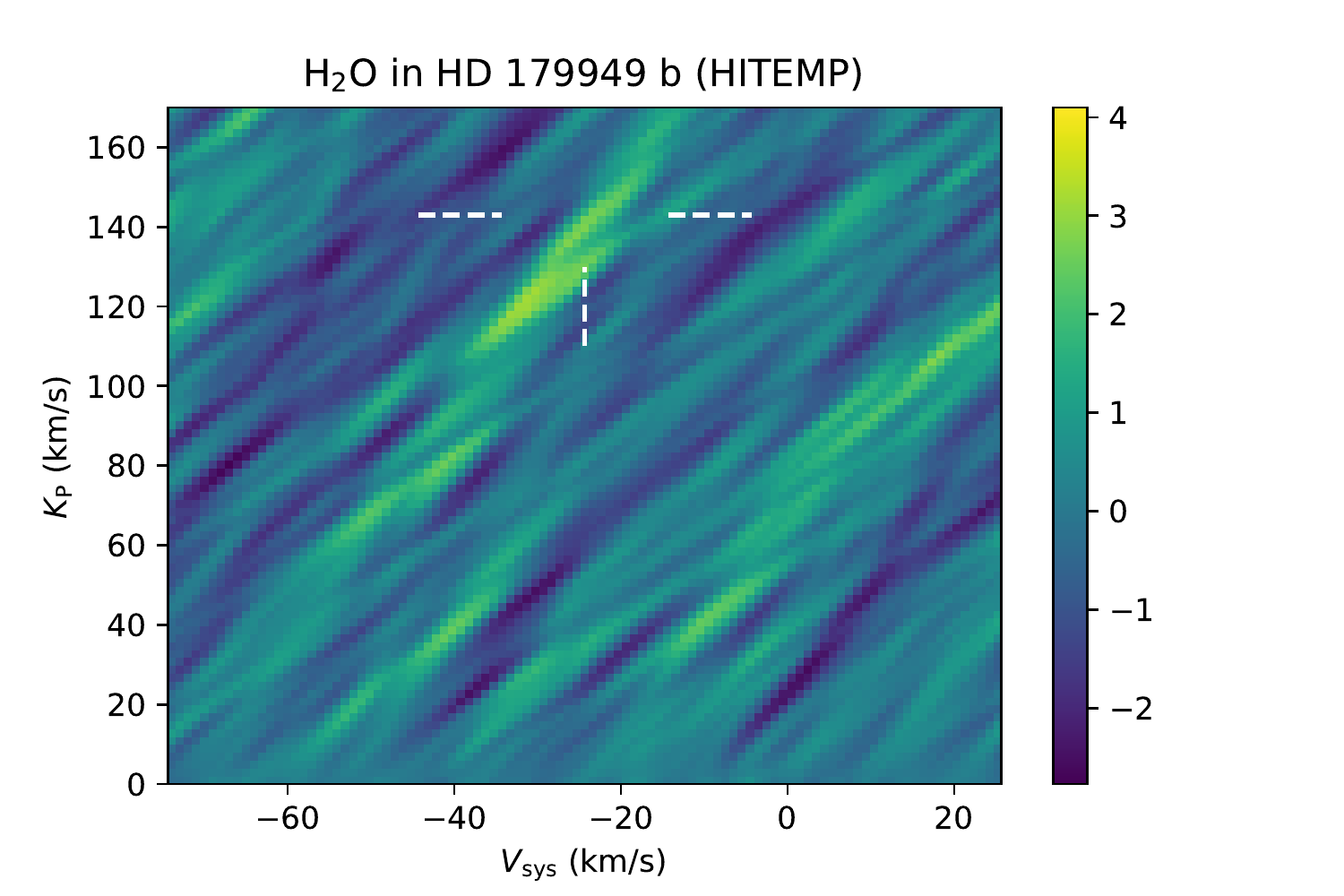}
	\includegraphics[width=0.42\textwidth,trim={1.8cm 0cm 2.0cm 0},clip]{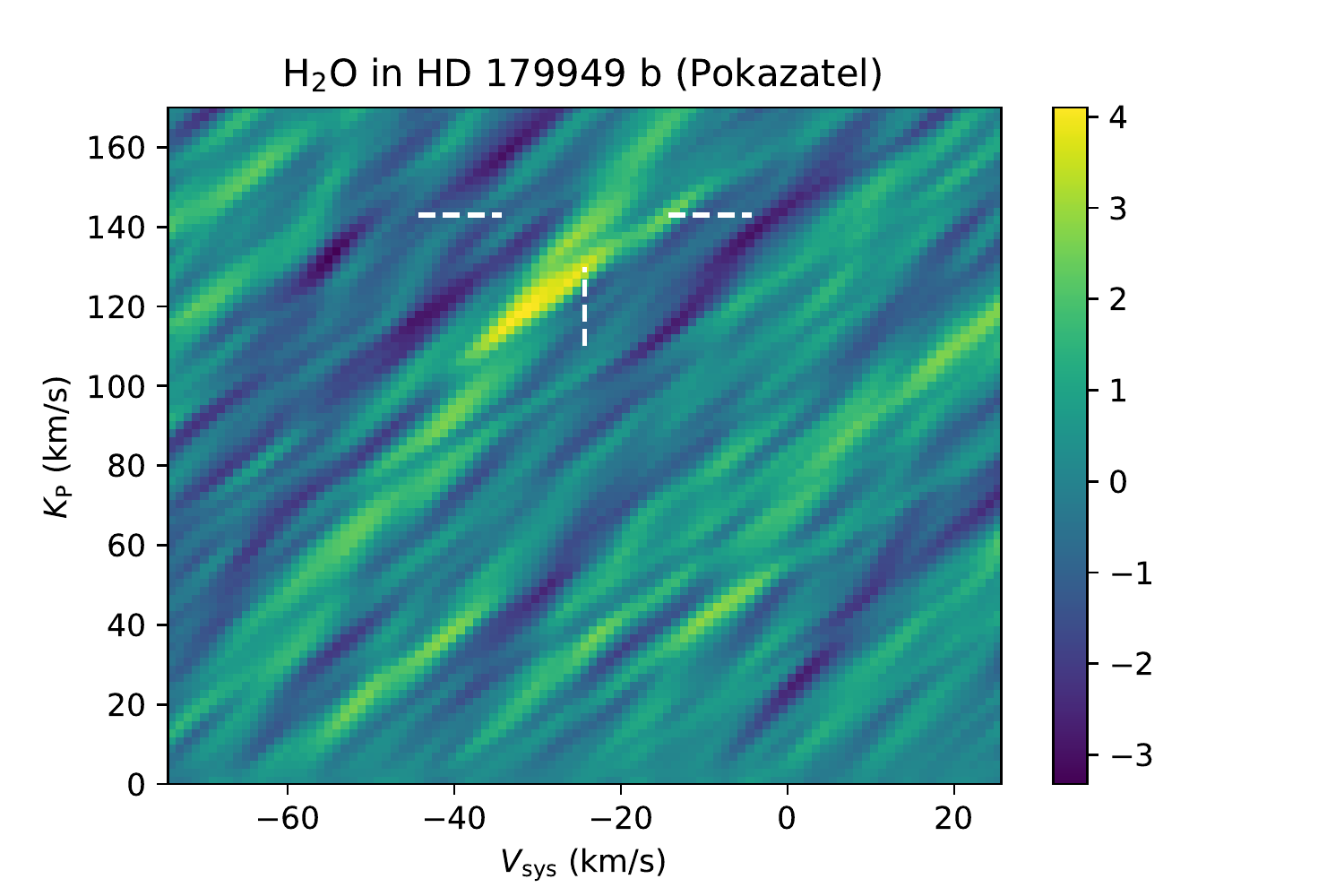}
    \caption{H$_2$O signal obtained by cross correlating VLT/CRIRES spectra of exoplanet HD~179949~b around 2.3 $\mu$m with a best-fitting model matching the parameters of \citet{brogi2014} and using either opacities from \citet{rothman2010} (left panel) or the newest opacities from \citet{polyansky2018} (right panel). The cross correlation signal is shown as function of systemic velocity $V_\mathrm{sys}$ and maximum orbital radial velocity $K_\mathrm{P}$. Although the two models show a signal in qualitative agreement, the best-fitting planet radial velocity semi-amplitude is in tension with the value obtained from the combined CO+H$_2$O analysis (dashed white lines).}
\label{fig:hd179949_ccf}
\end{figure*}

\section{Validation on Existing Observations of Hot Jupiters}\label{sec:prev_detections}

We now explore previous detections of molecular species with HRS in the atmospheres of hot Jupiters using the cross sections developed in section \ref{sec:cross_secs}. These hot Jupiter atmospheres have been extensively observed in a number of spectral ranges in the infrared at R$\gtrsim$50,000 from observational facilities such as CRIRES/VLT and GIANO/TNG. We discuss five hot Jupiters, HD~189733~b, HD~179949~b, $\tau$~Bo{\"o}tis~b, HD~209458~b and HD~102195~b, which have shown evidence for H$_2$O, CO, HCN and CH$_4$. We use the cross sections computed in section \ref{sec:cross_secs} to calculate the high resolution spectrum using GENESIS \citep{gandhi2017}. For each case we use the best fit parameters from each detection \citep{birkby2013, brogi2014, brogi2012, hawker2018, guilluy2019}. We then use these spectra to cross correlate against observations following the same procedure as discussed in previous work.

Unless explicitly specified, observations are processed as in the published detection papers, i.e. they use the same calibration and removal of telluric and stellar lines. The only difference is that we use the models described in this work to compute the cross correlation.

In what follows we adopt a threshold of S/N=3 to claim a tentative detection, and a S/N$>$4 to claim a definite detection. 

\subsection{\texorpdfstring{H$_2$O}{H2O}: HD~189733~b}\label{sec:h2o_hd189}

H$_2$O was first detected using low-resolution Spitzer data \citep{jt400} and confirmed with HRS in the dayside spectrum of HD~189733~b observed between 3.18 and 3.27~$\mu$m using the CRIRES spectrograph \citep{birkby2013}. Subsequent transmission spectra with GIANO \citep{brogi2018} and CARMENES \citep{alonso-floriano2019} have also shown evidence for H$_2$O in the atmosphere of this planet. We utilise the CRIRES observations from \citet{birkby2013}, calibrated as in their work for the alignment to the common reference frame of the Earth's atmosphere and the determination of the pixel-wavelength solution. Following the prescriptions of their work, we limit the analysis to detector 1 and 3 of CRIRES, given that detector 2 is affected by very low telluric transmission and detector 4 by a well-documented instrumental issue. We additionally also use the same mask as \citet{birkby2013} for our work. We use a more straightforward detrending algorithm than Sysrem to remove telluric and stellar lines based on singular-value decomposition (SVD), which does not allow for unequal error bars on the data points. We remove the same number of eigenvectors (9 for detector 1 and 3 for detector 3), and verify that as in the case of \citet{birkby2013} this is the optimal number of components to maximise the signal. The use of SVD over Sysrem increases the signal-to-noise ratio of the measured signals by about 0.7, regardless of the line list used to produce the model spectrum used for cross correlation.

As we show in Figure~\ref{fig:hd189733_ccf}, we recover a detection of water vapour at a S/N = 5.7 and at the expected planet position. We note that we obtain a comparable signal and very similar noise structure when we cross correlate with the best-fitting model of \citet{birkby2013}, which also contains water but from HITEMP \citep{rothman2010}. We therefore quantitatively confirm the result shown in Figure~\ref{fig:cs_h2o_compare} (right panel), and report a strong agreement between two different line lists for water around 3.2$\mu$m. 

\subsection{\texorpdfstring{H$_2$O}{H2O}: HD~179949~b}\label{sec:h2o_hd179}

Detection of water vapour in $K$-band CRIRES observations (2.27-2.35 $\mu$m) has been somewhat less convincing compared to the cases reported in Section~\ref{sec:h2o_hd189}. This is not completely unexpected, given that this spectral window is far from the main opacity peaks of water vapour (see Figure~\ref{fig:cs}), and CO is the main opacity source.
Consequently, water alone has produced either no signals \citep{brogi2012, brogi2017, brogi2019}, or signals that are just above the threshold of detectability \citep{brogi2014, gandhi2019b}, but still strengthening the cross correlation substantially when modelled in conjunction with CO. In this paper we re-examine one of the latter cases, namely 3 half-nights of dayside spectroscopy of the non-transiting planet HD~179949~b \citep{brogi2014}. Here water was detected at a S/N = 3.9, and with a noise pattern in velocity space marginally inconsistent with the detection coming from CO (S/N = 5.8). When combined, however, the cross correlation from the two species was shown to co-add constructively and deliver a convincing S/N = 6.3, or a significance of 5.8$\sigma$. 

We have modelled the atmosphere of HD~179949~b with a pressure-temperature ($P-T$) profile consistent with the family of best-fitting models in \citep{brogi2014}, and a water abundance of 10$^{-4.5}$. Once again, we have utilised cross sections computed from both HITEMP \citep{rothman2010} and POKAZATEL \citep{polyansky2018}. The resulting cross correlation is shown in Figure~\ref{fig:hd179949_ccf}. In spite of a qualitative agreement between the two signals and the overall noise structure, there is a net departure from the best-fitting velocities obtained from the combined H$_2$O+CO model. The cross correlation seems to be double-peaked, where the primary peak at lower $K_\mathrm{P}$ is just above the threshold for marginal detection (S/N = 3.2) with HITEMP, but peaks at S/N = 4.1 with POKAZATEL. Furthermore, the secondary peak matching the $K_\mathrm{P}$ of \citep{brogi2014} is marginally detected in both cases, albeit slightly more convincingly with the HITEMP line list. 

There are arguably astrophysical reasons that could produce a mismatch between velocities measured with two different species, e.g., strong atmospheric patterns variable with pressure. However, if we did not have any prior knowledge of the planet's orbit and only rely on water alone, we could be deriving a biased value of its radial velocity semi-amplitude, which consequently would affect both the inferred planet mass and orbital inclination. As the detection of water in HD~179949~b is just above marginal, we cannot firmly conclude that there is a potential influence of water line lists on results derived with $K$-band spectra. We point out instead that further investigation and possibly a larger amount of data is required to identify the possible reasons of this discrepancy.

\begin{figure}
\centering
	\includegraphics[width=0.49\textwidth,trim={0.2cm 0cm 1.5cm 0},clip]{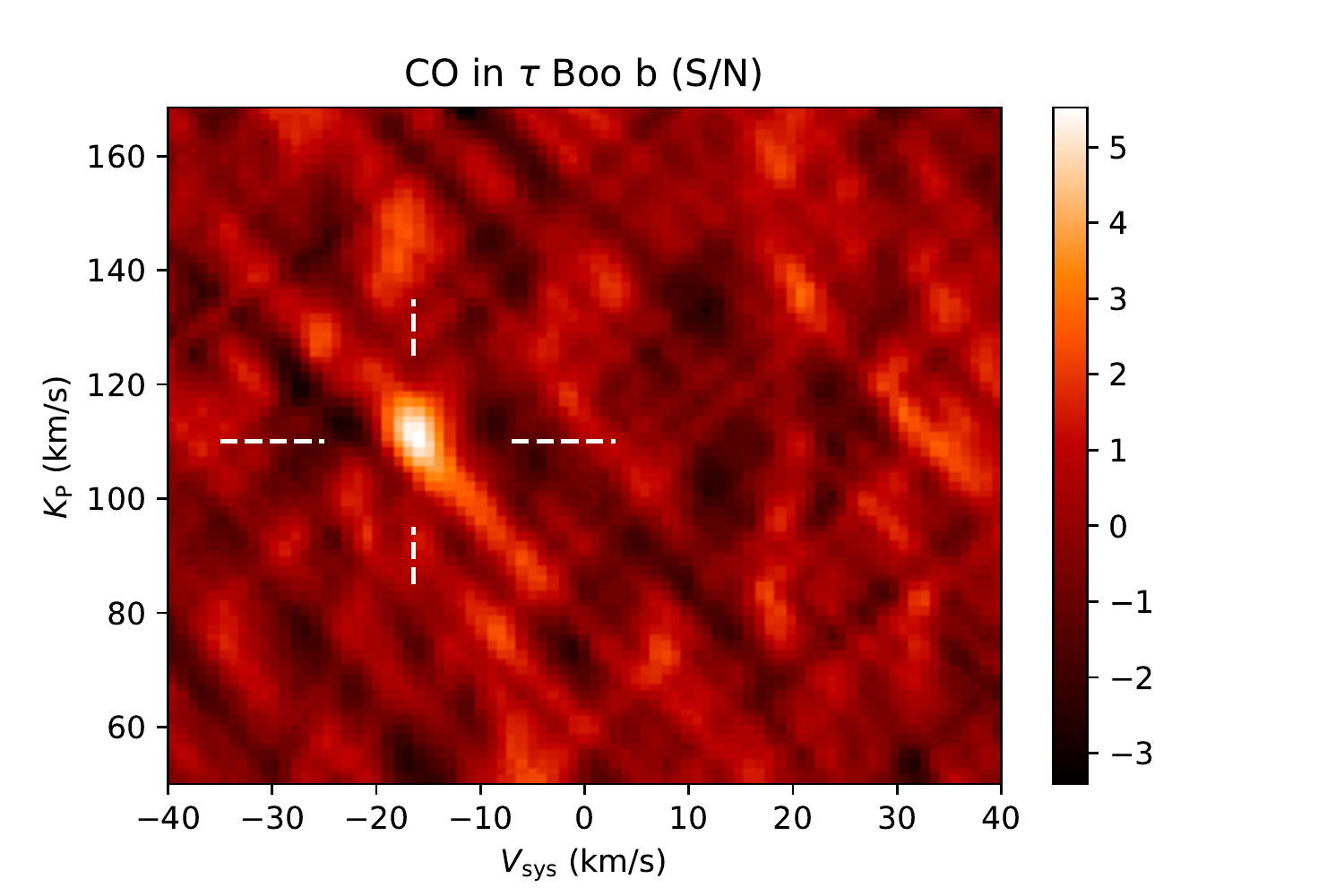}
	\vspace{-5mm}
    \caption{CO signal obtained by cross correlating VLT/CRIRES spectra of exoplanet $\tau$~Bo\"otis~b around 2.3 $\mu$m with a CO model matching the best-fit parameters of \citet{brogi2012}. The cross correlation signal is shown as function of systemic velocity $V_\mathrm{sys}$ and maximum orbital radial velocity $K_\mathrm{P}$. The known values of (-16.4, 110) km s$^{-1}$ are marked with dashed white lines and are a good match to these observations.}
\label{fig:tauboo_ccf}
\end{figure}

\subsection{CO: \texorpdfstring{$\tau$}{Tau}~Bo{\"o}tis~b}\label{sec:tauboo}

This dataset consists of three half nights of observations (approximately 3 $\times$ 6 hours including overheads) of the non-transiting planet $\tau$~Bo{\"o}tis~b. The spectra are observed with CRIRES at the Very Large Telescope and span the spectral range 2.29-2.35~$\mu$m, where CO possesses strong absorption (see Figure ~\ref{fig:cs}). As in \citet{brogi2012}, we exclude detectors 1 and 4 from the analysis and weigh the cross correlations from each spectrum and each detector equally. Results from the cross correlation analysis are presented in Figure~\ref{fig:tauboo_ccf}. We confirm the detection of \citet{brogi2012} at a compatible S/N of 5.6 and at the same values of planet systemic velocity ($V_\mathrm{sys} = -16.4$ km s$^{-1}$) and maximum orbital radial velocity ($K_\mathrm{P} = 110$ km s$^{-1}$). Consistently with \citet{brogi2012}, we find a marginal increase of the cross-correlation signal when including water as additional species. However, water-only models do not result in a detection above the threshold of S/N=3. Indeed, based on the increase in S/N with the mixed H$_2$O + CO model, we estimate that water vapour can only contribute by S/N$\le2$.

\subsection{HCN: HD~209458~b}

Evidence for HCN was first presented by \citet{hawker2018} based on the analysis of high resolution dayside observations of the hot Jupiter HD~209458~b obtained with CRIRES. We use their processed spectra for what concerns pixel-wavelength calibration, and removal of telluric lines through the Sysrem algorithm. We generate a model spectrum with the best fit parameters in \citet{hawker2018}, corresponding to an atmospheric abundance of 10$^{-5}$, and including $\gtrsim4\times10^5$ transitions in the observed spectral range (3.18-3.27~$\mu$m). 

\begin{figure}
\centering
	\includegraphics[width=0.49\textwidth,trim={0.2cm 0cm 1.5cm 0},clip]{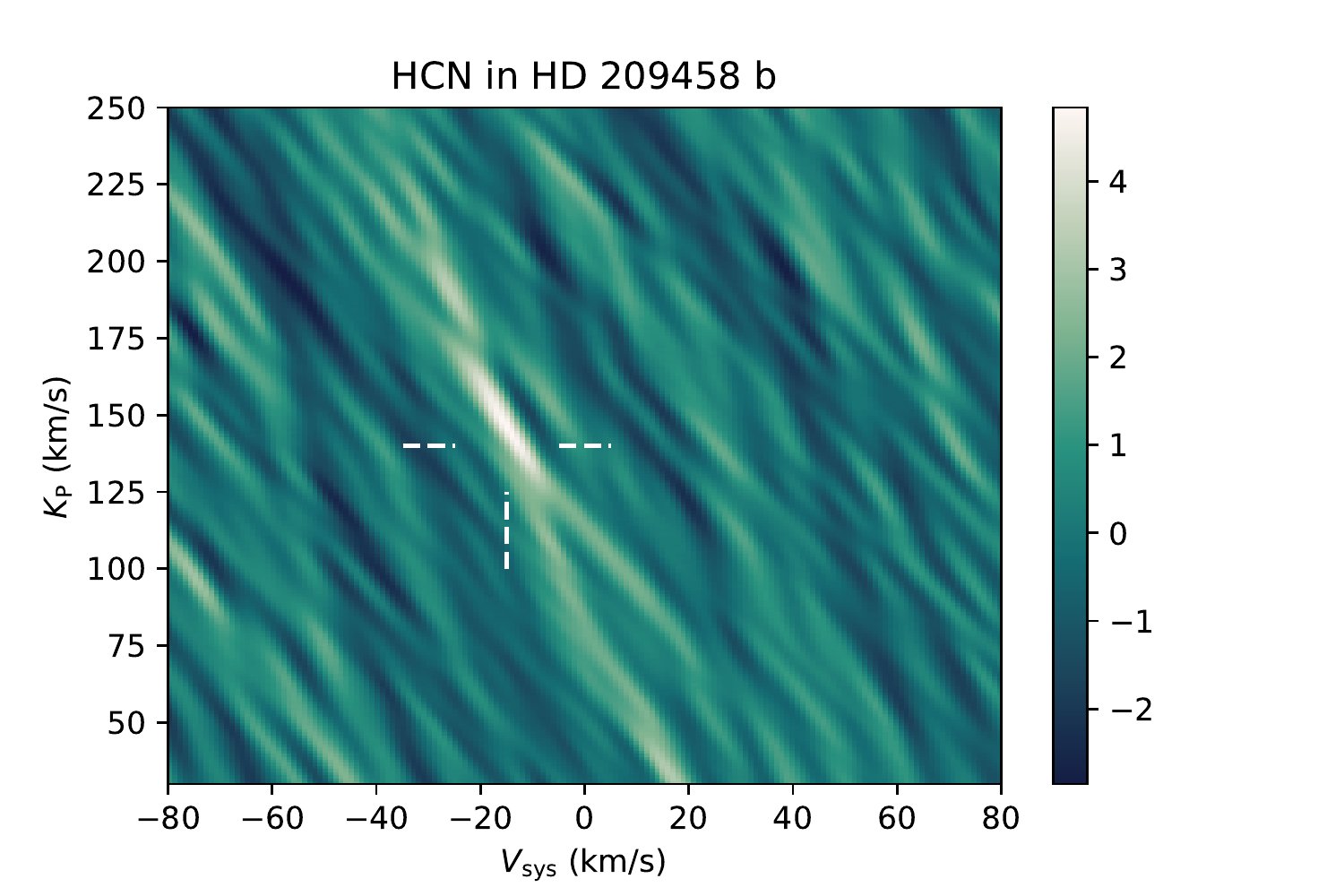}
    \caption{HCN signal obtained from cross correlation of L-band VLT/CRIRES spectra of exoplanet HD~209458~b with a best-fitting model matching the parameters of \citet{hawker2018}. We confirm a detection at a S/N = 4.8 and at the expected planet systemic and orbital radial velocity, indicated by the white dashed lines.}
\label{fig:hd209458b_ccf}
\end{figure}

As shown in figure \ref{fig:hd209458b_ccf}, we recover the detection of HCN at a S/N of 4.8 at the expected planetary radial velocity semi-amplitude and systemic velocity. The S/N of the recovered signal is fully consistent with previous work \citep{hawker2018} and the two dimensional velocity map (K$_p$ and V$_{sys}$ in the 3.18-3.27~$\mu$m range) also shows a very similar noise structure. This is in line with expectations, given that we use the same ExoMol line list, which is updated to include H$_2$ and He broadening. Our model thus has nearly identical line positions and strengths as the best fitting model of \citet{hawker2018}.  

\begin{figure*}
\centering
    \includegraphics[width=0.42\textwidth,trim={0.2cm 0cm 3.6cm 0cm}, clip]{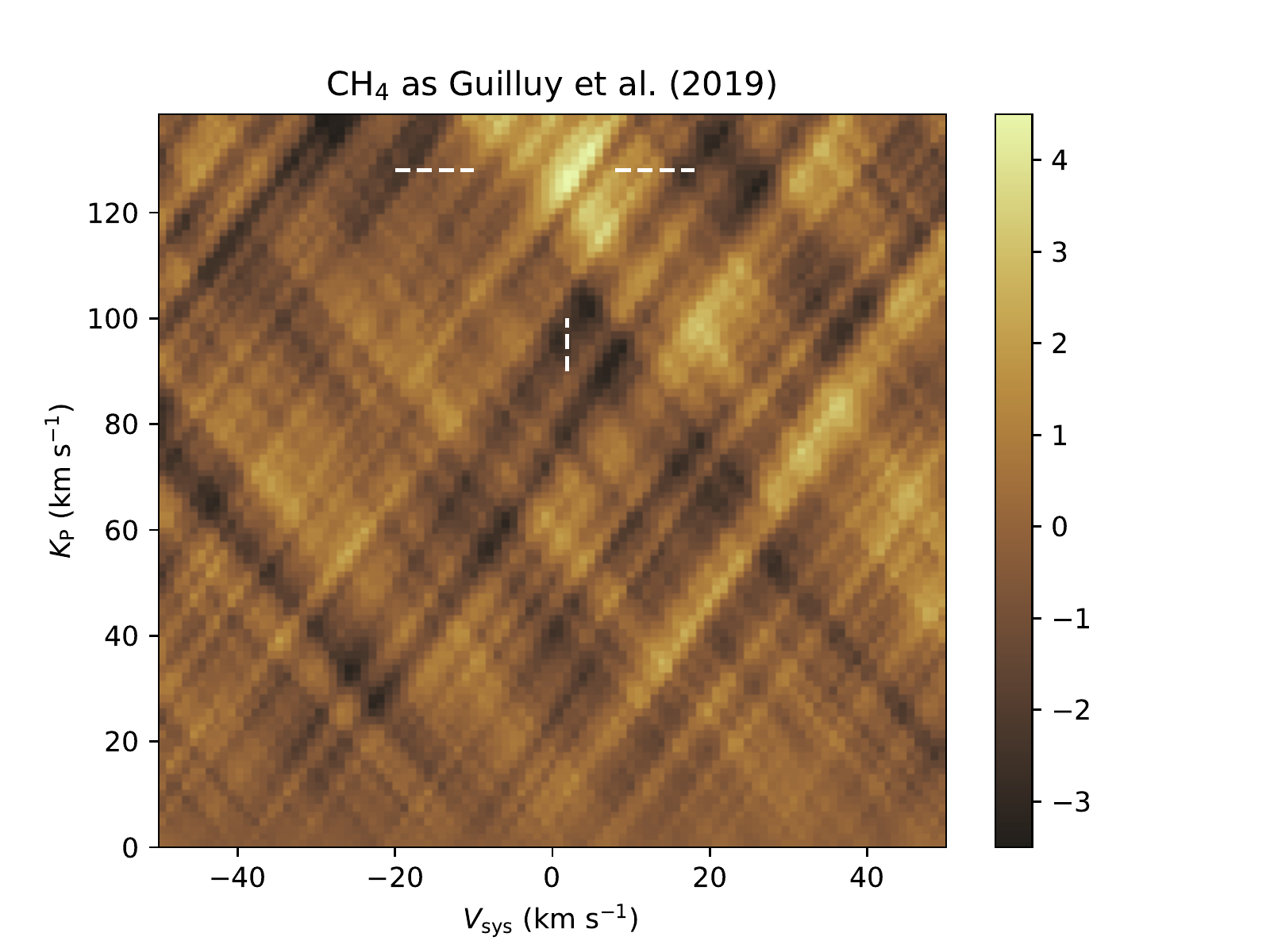}
    \includegraphics[width=0.42\textwidth,trim={1.8cm 0cm 2.0cm 0cm}, clip]{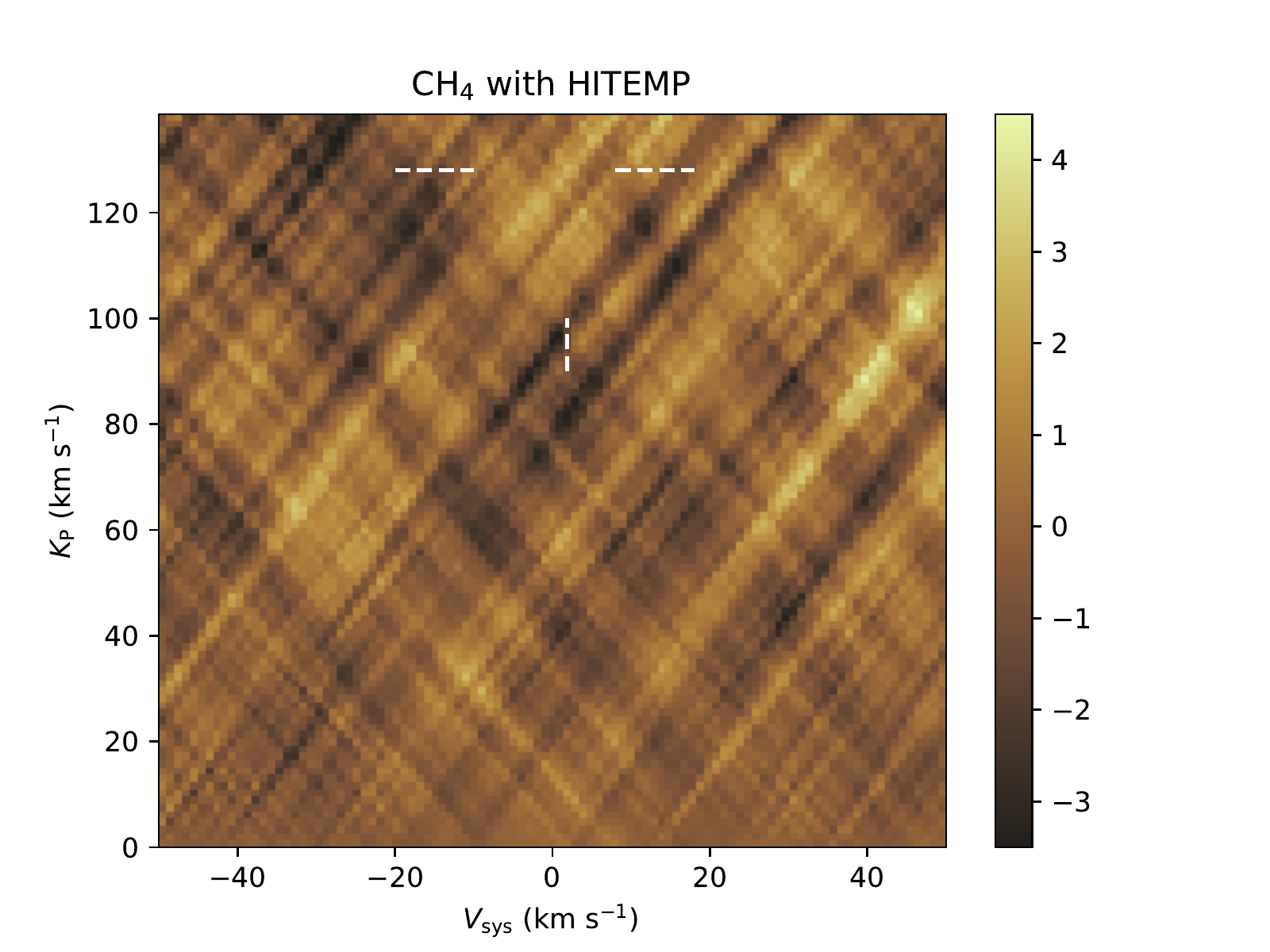}
    \caption{Cross correlation of models with methane as molecular species and dayside spectra of exoplanet HD~102195~b processed as in \citet{guilluy2019}. Left panel: cross correlation with models containing cross sections computed from HITRAN2016 \citep{gordon2017}. Right panel: cross correlation with a model containing the recent HITEMP line list \citep{hargreaves2020}. Cross correlation are given as function of systemic velocity and maximum orbital radial velocity.}
\label{fig:hd102195_ccf}
\end{figure*}

\subsection{\texorpdfstring{CH$_4$}{CH4}: HD~102195~b}

The first detection of methane at high spectral resolution was only recently reported by \citet{guilluy2019}, who analysed three half-nights of dayside spectra of the non transiting planet HD~102195~b observed with GIANO at the Telescopio Nazionale Galileo. The spectra span orbital phases prior, around, and post superior conjunction, and this sampling was adopted to obtain the tightest constraint on the orbital parameters of the exoplanet, as done in the case of $\tau$~Bo\"otis~b. In this work for full consistency we utilise the data processed and masked as in \citet{guilluy2019}, resulting in coverage of most the $K$ band (2.1-2.4~$\micron$ split in 4 or 5 spectral orders), part of the $H$-band (2-3 orders in the range 1.6-1.8~$\micron$) and part of the J band (2 orders between 1.0-1.2~$\micron$) at a resolving power of 50,000. We simultaneously use all of the available orders across all three bands for our analysis. This choice corresponds to 9 or 10 orders (variable with the observing night) out of 50, and matched the choices in \citet{guilluy2019}. The incomplete coverage results from combination of a challenging wavelength calibration of some of the spectral orders, and the choice to only include the orders corresponding to the highest methane cross section, as shown for instance in Figure~\ref{fig:cs}. 

In Figure~\ref{fig:hd102195_ccf} we present the total cross correlation signal coming from two different models. The former, shown on the left panel, is obtained with cross section computed from the full list of lines published by the HITRAN group \citep{gordon2017} at the time of the publication of \citet{guilluy2019}. The total cross correlation signal is comparable to their work in terms of both strength (S/N = 4.6) and position in velocity space. The latter model, shown on the right panel, utilises the cross sections described in this paper and obtained from the HITEMP database \citep{hargreaves2020}. This model shows no significant correlation at the expected position of the planet (marked by dashed lines), thus we are unable to confirm the detection with this more recent line list.

In interpreting this result, the following elements need to be considered. Firstly, we note that the detection of \citet{guilluy2019} is substantiated by the simultaneous measurement of water vapour at compatible velocities. This allows the signal of the two molecular species to coherently combine when a model containing CH$_4$+H$_2$O is used for cross correlation. We are indeed able to reproduce the detection of water vapour in this study. Secondly, by looking at the right panel of Figure~\ref{fig:hd102195_ccf}, we also note that from data spanning a large range in orbital phase we would expect a much tighter localisation of the cross correlation peak, e.g. the spot-like detection observed from CO in $\tau$~Bo\"otis~b (see section~\ref{sec:tauboo}). The qualitative difference in the shape of the cross-correlation peak could be due to the fact that this non-transiting exoplanet has a poorly constrained orbital solution. By propagating the error bars in orbital period and time of inferior conjunction, we estimate an uncertainty of about 10\% in orbital phase. 

\begin{figure}
\centering
	\includegraphics[width=0.49\textwidth,trim={0.5cm 0cm 0.5cm  0},clip]{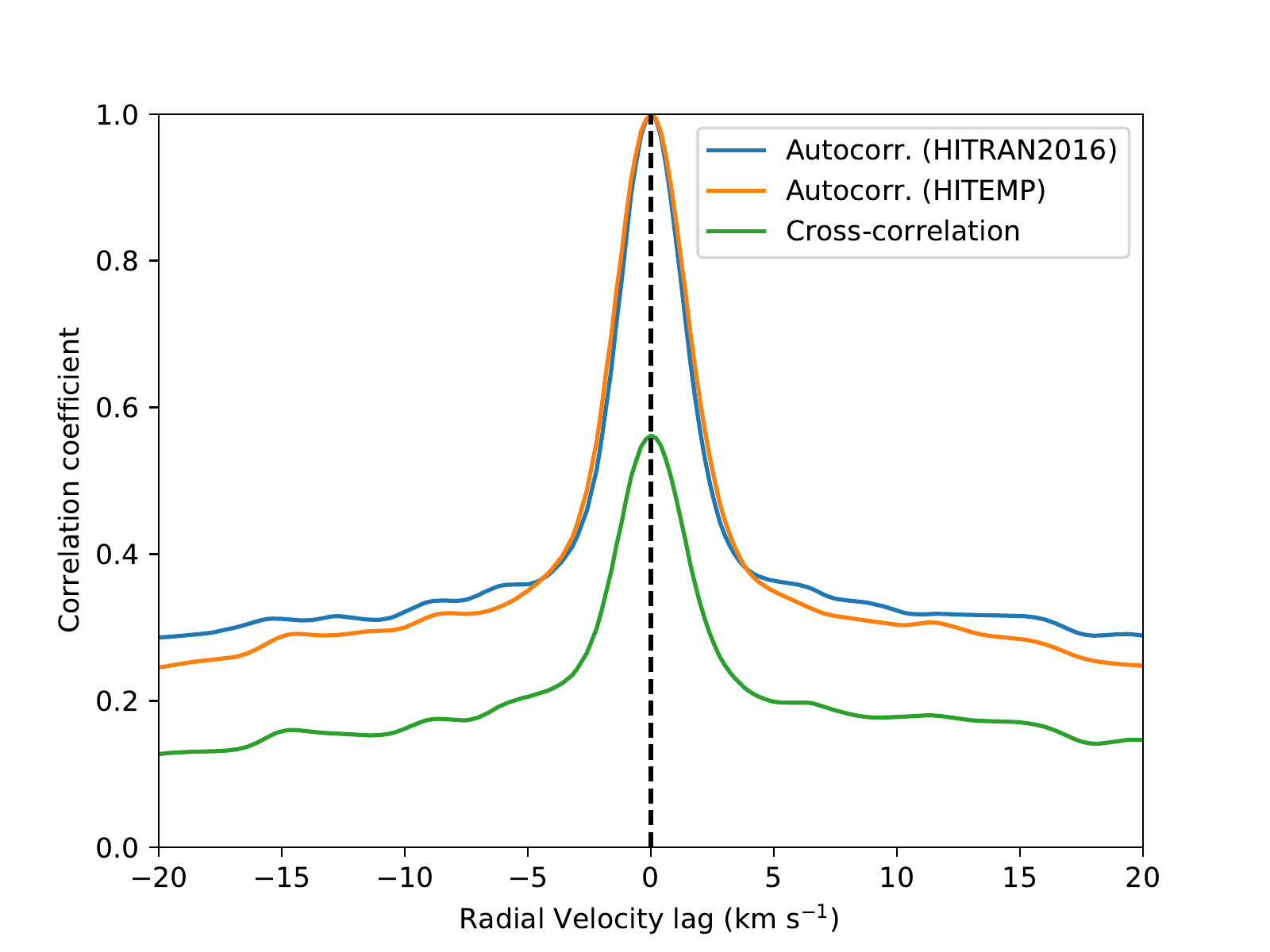}
    \caption{Auto-correlation function of the model computed with HITRAN2016 (blue line), HITEMP (orange line), and the cross correlation function of the two models (green line). The latter peaks at 0.56, highlighting substantial differences in the spectral features contained by the two models.}
\label{fig:model_autocorrelation}
\end{figure}

Thirdly, we highlight significant differences between the spectra produced with HITRAN2016 and HITEMP. While all the strong spectral lines agree well between the two spectra, the overall continuum level does not, and this is expected because HITEMP is much more complete at high temperatures, therefore raising the overall opacity due to methane. This increased opacity mutes some of the lines that in the HITRAN spectrum would appear as strong lines, and add a dense forest of weaker lines that were not present at all, therefore potentially affecting the cross correlation signal if these weaker lines carry a non-negligible weight. In order to assess their importance we show the auto-correlation of each of the models in Figure~\ref{fig:model_autocorrelation}, compared to their mutual cross-correlation signal. The latter peaks at a correlation value of 0.56, which indicates that a non-negligible fraction of the signal is carried by the weak lines present in HITEMP, but absent in HITRAN2016. As a consequence, if neglected when interpreting observations, the resulting cross correlation signal could be significantly muted. On the other hand, if these lines are included in the models but suffer from inaccuracies at the resolving power of the observations, this could also hamper the methane signal preventing detection. 

The assumed P-T profile of the atmosphere may also more strongly impact the HITEMP model spectrum. The weaker spectral lines are more numerous and generally vary more significantly with temperature compared to the strongest lines. Thus if the overall atmospheric temperature is less than that assumed in the model, the more complete HITEMP line list may perform worse as the many more weak lines contribute to reduce the overall correlation with the data. We tested this by cross correlating the data with models of HD~102195~b generated with a P-T profile that was 300~K cooler. This was the coolest P-T profile that remained consistent with the expected physical conditions from the orbit of the planet. With this cooler temperature profile the cross correlation between the HITEMP and HITRAN2016 models did improve to $\sim$0.62 but we found no significant improvement in the overall signal with the HITEMP line list. This shows that while the contribution from the weak lines present in HITEMP reduces as we approach cooler temperatures, this change is not significant enough to explain the current differences in our findings. 

We also note that the CH$_4$ cross section is most accurately determined at low wavenumbers \citep{hargreaves2020} and this may also influence the overall detection if it is dominated by the higher frequency bands such as the H band. Based on the limited data presented here and the inaccuracy of the orbital solution, in this context we limit to report these caveats and we conclude that it is still challenging to derive firm prescriptions on the methane cross sections without a dedicated follow-up study.

\section{Discussion and Conclusions}\label{sec:conclusion}

We present new publicly available high resolution cross sections\footnote{The cross sections can be accessed from the Open Science Framework via the following link: \url{ https://osf.io/mgnw5/?view_only=5d58b814328e4600862ccfae4720acc3}} for the volatile molecular species expected to be prominent in hot Jupiters, warm Neptunes and super Earths with temperatures ranging from $\sim$500-1500~K. We focus on species with observable signatures from current ground based facilities in the near infrared. This leads us to consider line lists sources for six species, namely H$_2$O, CO, HCN, CH$_4$, NH$_3$ and CO$_2$, which are given in Table~\ref{tab:line_list_table}. We compute the cross sections for these species in the wavelength range 0.95-5~$\mu$m (2000-10,526~cm$^{-1}$) at a wavenumber spacing of 0.01~cm$^{-1}$. This corresponds to a spectral resolution of $\mathrm{R}=10^6$ at 1~$\mu$m. We summarise the key developments of the cross sections below.

\begin{itemize}
    \item The line list sources for the cross sections are given in Table~\ref{tab:line_list_table} and use the ExoMol \citep{tennyson2016}, HITEMP \citep{rothman2010, hargreaves2020} and Ames \citep{huang2017} databases.
    
    \item These line lists were chosen as they typically use ab initio calculations combined with experimentally verified line positions for maximum accuracy in the line position. This is key for detections with HRS in order to maximise the cross correlation with the observations.
    
    \item The completeness of the line lists in the spectral range provided also ensures that the high resolution cross sections computed in this work are optimised for use in low resolution spectra. 
    
    \item Each line is broadened according to the pressure and temperature from the grid given in Table~\ref{table:cross_sec_grid}. This grid of $P-T$ values was chosen to represent typical photosphere conditions of super Earths, warm Neptunes and hot Jupiters. The overall line profile of each transition in a line list is a Voigt function resulting from a convolution of a Gaussian from thermal broadening and a Lorentzian from pressure broadening. In addition, natural broadening has been included which occurs from the uncertainty principle which also results in a Lorentzian line shape \citep{gray1976}.
    
    \item We use the latest H$_2$ and He pressure broadening for the cross sections of each species thanks to recent work on pressure broadening coefficients (see Table~\ref{tab:line_list_table}). This allows for more accurate cross sections of these planets at high pressure. These pressure broadening coefficients are particularly beneficial for accurate low resolution spectra which are more sensitive to the line wings where this is the dominant source of broadening \citep{hedges2016}.
    
    \item As more accurate and complete line lists become available in the near future we will continually update our cross sections to ensure that the most up to date are available.
\end{itemize}

We use the cross sections calculated in this work to generate spectra of known planets to identify spectral features of each species in the infrared. We model the atmosphere of three exoplanets, the super Earth/sub-Neptune GJ~1214~b, the warm sub-Neptune GJ~3470~b and the hot Jupiter HD~189733~b, chosen to have a wide range in mass, radii and equilibrium temperatures. We find that H$_2$O has prominent spectral features throughout the infrared for each of the exoplanets modelled given its abundance and strong cross section. CO on the other hand only has strong features for hot Jupiters such as HD~189733~b where it is expected to be more abundant \citep[e.g.][]{madhu2012}. The abundance and hence spectral features of other species such as HCN can be strongly affected by the atmospheric C/O ratio in hot Jupiters \citep{moses2013, drummond2019}. The cooler planets GJ~1214~b and GJ~436~b show prominent spectral features for CH$_4$ and NH$_3$ given the higher abundance for these two planets. In particular, the strongest features for these species occur in the wavelength ranges where H$_2$O absorption is weak. This thus represents an exciting opportunity in the future for chemical detections using HRS which typically probe in between H$_2$O bands where many of these features may be detectable.

We also use our new cross sections to generate high resolution spectra of hot Jupiters which have shown evidence for these volatile molecular species from previous observations \citep{birkby2013, brogi2014, brogi2012, hawker2018, guilluy2019} to validate our opacities. We model the best fitting spectra of HD~189733~b and HD~179949~b with H$_2$O, $\tau$~Bo{\"o}tis~b with CO, HD~209458~b with HCN and HD~102195~b with CH$_4$ using GENESIS \citep{gandhi2017}. We then perform cross correlation on the observations following similar data analysis techniques discussed in previous works. We reproduce the high signal to noise detections seen for H$_2$O, CO and HCN at the previously measured values of planetary and systemic velocity for each planet which had shown a clear detection. HD~179949~b on the other hand showed only weak constraints on H$_2$O from the $K$-band observations, but this is also consistent with that reported in \citet{brogi2014} previously. In addition, the CH$_4$ detection could be reproduced in HD~102195~b with the HITRAN line list \citep{gordon2017} used in \citet{guilluy2019}, but not with the new HITEMP line list \citep{hargreaves2020}. While in this work we only report the discrepancy between the result coming from two line lists, we suggest that further observations of a wider range of exoplanet atmospheres are needed to assess the current state of the art of methane line lists for high resolution spectroscopy of exoplanets. 

In the next few years instruments such as SPIRou \citep{artigau2014}, GIANO \citep{oliva2006}, CARMENES \citep{quirrenbach2014}, IGRINS \citep{park2014}, NIRSPEC \citep{mclean1998}, iSHELL \citep{rayner2016} and CRIRES+ \citep{follert2014} will provide detections of numerous molecular species in the atmospheres of transiting and non-transiting exoplanets in the infrared. In addition, high resolution retrievals of such observations are now becoming possible and are capable of statistically robust abundance estimates and detection significances, with precision similar to space based lower resolution observations \citep{brogi2017, brogi2019, gandhi2019b}. Hence such high resolution opacities are crucial for abundance estimates. These will also be particularly important as we explore atmospheric chemistry on cooler and smaller exoplanets given that HRS is sensitive to trace species in the atmosphere \citep{snellen2010, birkby2018}. In the future large ground based telescopes such as the ELT, TMT and GMT may also be the most viable means to search for biosignatures on small rocky planets \citep{snellen2013, rodler2014, hawker2019, lopez-morales2019}. Accurate spectra and hence accurate cross sections of these species are therefore vital for atmospheric characterisation with HRS.

\section*{Acknowledgements}

SG and MB acknowledge support from the UK Science and Technology Facilities Council (STFC) research grant ST/S000631/1; SY and JT acknowledge support from STFC grant ST/R000476/1; PC was supported by the UK Engineering and Physical Sciences Research Council (EPSRC) grant EP/M506448/1 and Servomex Ltd. JLB acknowledges funding from the European Research Council (ERC) under the European Union's Horizon 2020 research and innovation program under grant agreement No 805445. G.G. acknowledges the financial support of the 2017 PhD fellowship programme of INAF. This work makes use of observations made using the CRIRES spectrograph on the European Southern Observatory (ESO) Very Large Telescope (VLT). We thank the ESO Science Archive for providing the data. We would like to thank Robert Hargreaves for providing the HITEMP CH$_4$ line list. We also thank the anonymous referee for a careful review of the manuscript.




\bibliographystyle{mnras}
\bibliography{references} 




\bsp	
\label{lastpage}
\end{document}